# Resolving shortwave and longwave irradiation distributions across the human body in outdoor built environments

*Kambiz Sadeghi,[1,2] Shri H. Viswanathan,[1] Ankit Joshi,[1,2] Lyle Bartels,[1] Sylwester Wereski,[3] Cibin T. Jose,[1] Galina Mihaleva,[4] Muhammad Abdullah,[5] Ariane Middel,[2,6,7] and Konrad Rykaczewski,[1,2*]*

1. *School for Engineering of Matter, Transport and Energy, Arizona State University, Tempe, AZ 85282, USA*
2. *Julie Ann Wrigley Global Futures Laboratory, Arizona State University, Tempe, AZ 85282, USA*
3. *Institute of Earth and Environmental Sciences, Maria Curie-Skłodowska University, Lublin, Poland*
4. *Arizona State University Fashion Institute of Design and Merchandizing (ASU FIDM), Herberger Institute for Design and the Arts, Arizona State University, Tempe, AZ 85282, USA*
5. *School of Sustainability, Arizona State University, Tempe, AZ 85282, USA*
6. *School of Arts, Media and Engineering, Herberger Institute for Design and the Arts, Arizona State University, Tempe, AZ 85282, USA*
7. *School of Computing and Augmented Intelligence, Arizona State University, Tempe, AZ 85282, USA*

**corresponding author email: konradr@asu.edu, phone: (480) 965-4912, address: Arizona State University, 501 E Tyler Mall, Tempe, AZ 85282, USA*

**Abstract:** Outdoor built environments can be designed to enhance thermal comfort, yet the relationship between the two is often assessed in whole-body terms, overlooking the asymmetric nature of thermal interactions between the human body and its surroundings. Moreover, the radiative component of heat exchange—dominant in hot and dry climates—is typically lumped into a single artificial metric, the mean radiant temperature, rather than being resolved into its shortwave and longwave spectral components. The shortwave irradiation distribution on the human body is often highly anisotropic, causing localized thermal discomfort in outdoor environments. However, no existing methods effectively quantify shortwave and longwave irradiation distributions on the human body. To address this gap, we developed two methods to quantify these processes. The first approach uses an outdoor thermal manikin with a white-coated side, enabling the separation of spectral components by subtracting measurements from symmetrically corresponding surface zones of tan color. The second hybrid approach converts radiometer measurements in six directions into boundary conditions for computational thermal manikin simulations. We evaluated irradiation distributions for various body parts using both methods during outdoor measurements across sunny, partially shaded, and fully shaded sites under warm to extremely hot conditions. In most cases, the two methods produced closely aligned results, with divergences highlighting their respective strengths and limitations. Additionally, we used the manikin to quantify irradiation attenuation provided by five long-sleeve shirts with colors ranging from white to black. These advanced methods can be integrated with airflow and thermoregulatory modeling to optimize outdoor built environments for enhanced human thermal comfort.

**Highlights:**

- Two-color thermal manikin measures shortwave and longwave radiation on its body parts
- Simulation input model based on 6-directional radiometer measurements is formulated
- Radiation simulation of a virtual manikin twin yields exposure across its body parts
- Methods align in outdoor measurements across sunny and fully shaded sites
- Manikin quantified irradiation attenuation by five long-sleeve shirts from white to black

**Keywords:** outdoor radiation exchange, shortwave radiation, longwave radiation, outdoor thermal manikin, computational thermal manikin, integrated radiation measurement, clothing

## 1. Introduction

The increasing frequency and intensity of heat waves [1,2] and associated negative impacts on health [3,4], livability [5], and productivity [6–8] across the globe are stimulating efforts to design outdoor built environments to promote thermal comfort and safety [9–20]. The relationship between the outdoor built environment and thermal comfort [11,12,17,18,21–31] can be assessed in many ways, but most commonly it is related using empirical indices [32]. However, many of these indices were developed for indoor conditions that can be assumed to be uniform and steady, and despite some modifications for outdoor, are often poor predictors of comfort in outdoor settings that are characterized by transient and non-uniform conditions [24–26,33–37] and where air turbulence [38–45] and anisotropic radiation play important roles [34,46–54]. The radiative component of human-surrounding heat exchange—dominant in hot, dry climates [54,55]—is typically lumped into a single artificial metric, the mean radiant temperature (MRT)[56–58]. This parameter is often measured using globe [59–61] and cylindrical radiometers [62,63]. However, these affordable devices were developed for indoor settings [60] and are prone to significant MRT measurement errors in outdoor conditions [62], mandating empirical corrections [59,62] or simultaneous use of multiple globes or cylinders with varied colors and internal heating [61,63].The additional advantage of the latter multi-device radiometers is that beyond measuring MRT, they can also resolve the radiation absorbed by the human body into its shortwave (solar) and longwave (infrared) spectral components. The directional variation of these spectral components can be quantified using more advanced methods, such as arrays of multiple pyrgeometers and pyranometers [57,59,64,65] or panoramic spectral imaging [66–68].

Resolving the directional and spectral components of the radiation field is important since shortwave, and in some cases longwave, irradiation is highly anisotropic, which leads to uneven distribution patterns across the human body that can cause localized thermal discomfort [54,69–73]. The direct solar beam and complex shortwave interactions with the built environment impact this factor. For example, well-intentioned modifications to the built environment such as "cool pavements", which decrease the shortwave energy absorbed by the ground [74–77], can increase the exposure of a pedestrian to shortwave and total radiation [78,79], negatively impacting thermal comfort [80,81]. Importantly, an individual can easily attenuate their shortwave exposure using, for example, a parasol [82] or by wearing light color or radiatively-engineered clothing [83–85], a factor that would not be reflected in MRT since it assumes an "average" shortwave absorptivity of 0.7 [56]. The longwave radiation that a person is exposed to can also be anisotropic outdoors with, for example, sun-heated surfaces emanating higher fluxes or cold panels providing radiative cooling [86,87]. However, these variations are generally more minor than in the case of shortwave radiation [63,88]. In either case, quantifying outdoor anisotropic radiation distribution on the human body is a complex challenge.

Human exposure to anisotropic radiation has been quantified using tabulated view factors, built environment-focused simulations, and human-centric simulations. The anisotropic irradiation flux from a limited number of radiant heat sources or sinks is traditionally simplified into a plane radiant temperature or incorporated into energy balance calculations using tabulated view factors [54,71]. View factors have been quantified for seated and standing young adults [47,71], varied body types including children, and during motion [48,89–92]. They also have been applied to estimate MRT for thermal comfort calculations in indoor and outdoor settings [48,89,90,93]. This approach becomes cumbersome in complex urban surroundings, for which highly anisotropic shortwave and longwave radiative fluxes are commonly predicted using built environment-focused simulation tools such as RayMan [94], SOLWEIG [95], and ENVI-met [96] that consider local urban design and weather factors. However, these programs calculate the integrated exposure onto a point, sphere, cube, or cylinder; therefore, they cannot resolve exposure across different human body parts. Similarly, even advanced multidirectional anisotropic radiation measurements based on radiometer arrays [57,59,64,65] or panoramic spectral imaging [66–68] assume the human body to be a point, sphere, or cylinder. A realistic three-dimensional body shape is the basis of human-centric radiation simulations; however, these have predominantly investigated indoor settings [53,60,97–108], with only a few recent efforts focused on outdoor locations [86,88]. Critically, currently, there are no experiments quantifying radiative fluxes across the human body for benchmarking such simulations (i.e., the few notable efforts measured human skin temperature [46,50–52,69,86], which only estimates total radiative heat flux and requires additional calculations and assumptions that can introduce significant errors [109,110]).

To address this gap, we developed two methods to quantify shortwave and longwave irradiation fluxes across multiple human body parts in outdoor settings. The first approach uses the one-of-a-kind outdoor and extreme heat-compatible thermal manikin "ANDI" (see **Fig.1a-c**), which we recently demonstrated can be used to measure the total radiative heat flux absorbed across its parts in outdoor sunny and hot conditions [43]. Here, we introduce the "two-color" manikin method to separate the spectral-band irradiation components. The second method is hybrid, converting experimental irradiation measurements in 6-direction obtained using "MaRTy" biometeorological cart [64] (see **Fig.1d**) into boundary conditions [88] for computational simulations (see **Fig.1e**) employing ANDI's virtual twin (see **Fig.1f**). We evaluated irradiation distributions for various body parts using both methods during outdoor measurements across sunny, partially shaded, and fully shaded sites under warm to extremely hot conditions (see **Fig. 2**). Additionally, we used the manikin to quantify irradiation attenuation provided by five long-sleeve shirts in colors ranging from white to black. The specifics of the methods and measurements are described next, and are followed by the investigation outcomes and their discussion.

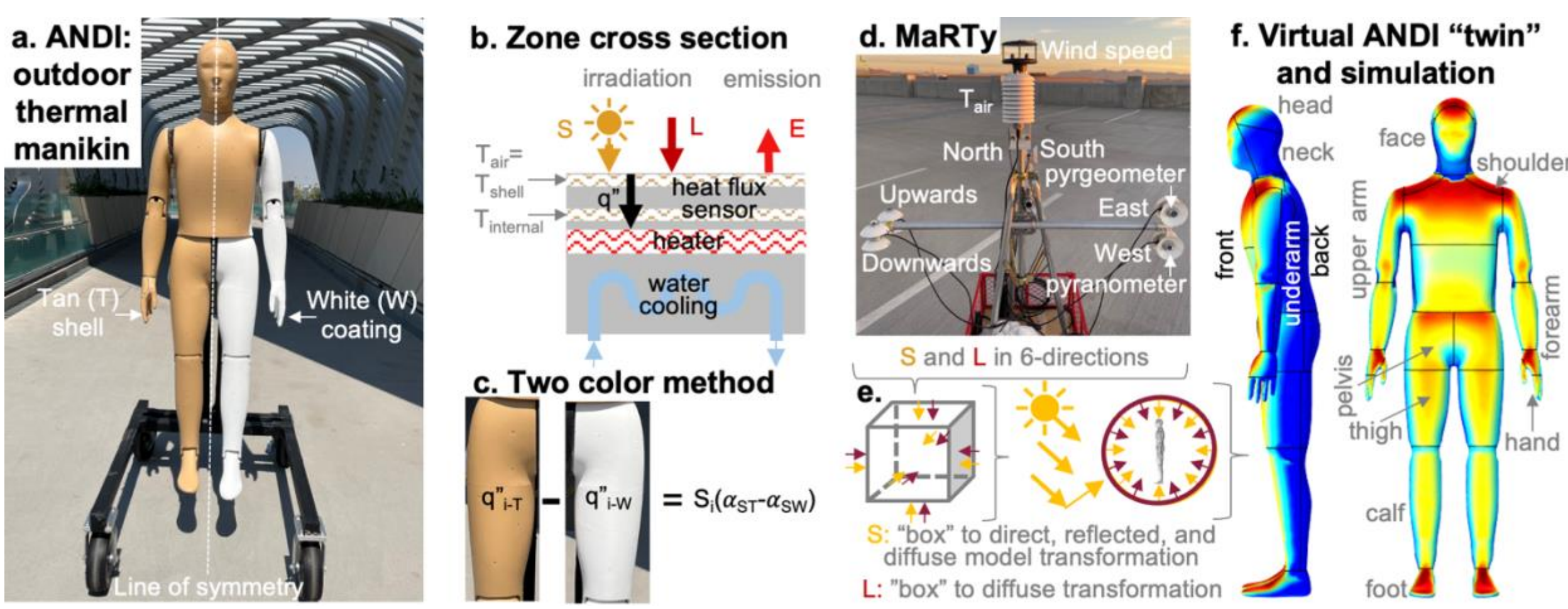

**Fig.1 Two methods to resolve shortwave and longwave irradiation distributions across the human body: a.** image of the outdoor thermal manikin ANDI with one set of symmetric zones with a white coating

(referred to with subscript W with corresponding shortwave absorptivity $\alpha_{SW}$) over the original tan (referred to with subscript T with corresponding shortwave absorptivity $\alpha_{ST}$) shell that is visible elsewhere, **b.** schematic of a cross-section of one of ANDI's internally cooled zones showing shortwave ($S$) and longwave ($L$) irradiation and longwave emitted fluxes ($E$) whose sum leads to the measured net heat flux within the shell ($q^{"}$); air, external shell (that is set to air one), and internal temperature are also shown **c.** illustration of the two-color ANDI method for calculating the shortwave irradiation on zone "i" ($S_i$) based on the subtraction of the heat flux for the white zone ($q^{"}_{i-W}$) from its corresponding tan zone one ($q^{"}_{i-T}$); **d.** image of MaRTy biometeorological cart used to collect short and longwave fluxes in six directions (east, west, south, north, upwards, downwards), which **e.** shows are transformed from "box" to direct, reflected, diffuse shortwave and diffuse longwave boundary conditions for **f.** virtual ANDI "twin" simulations with names of the zones indicated.

## 2. Methods

### *2.1 Overview*

We developed two approaches to quantify the shortwave and longwave irradiation across the human body. In the first based approach, we eliminated convective effects by setting the manikin shell temperature equal to that of air, measuring the absorbed heat flux using in-shell sensors (see **Fig.1b-c**), and accounting for emitted radiation [43]. We separate the spectral-band irradiation components by subtracting heat fluxes measured on zones with white and tan coatings. We developed two variants of this two-color ANDI method. One variant of this method relies on fluxes measured simultaneously on a zone covered with white paint and a symmetric one across the sagittal plane that is tan colored (see **Fig.1a**). The second variant of this method relies on sequential heat flux measurements with the "bare" (tan) manikin and with the manikin dressed in a whole-body white suit. The second method converts MaRTy's 6-directional shortwave and longwave measurements into boundary conditions [88] (see **Fig.1e**) for simulations employing ANDI's virtual twin (see **Fig.1f**). The following sub-sections describe the locations and types of field measurements (see **Fig. 2**) and details of the two approaches to quantify irradiation on the human body.

### *2.2 Locations and types of conducted field experiments*

We conducted three types of experiments with side-by-side ANDI and MaRTy, including "extended exposure" measurements with full variation of zenith angle from about solar noon to sunset in a mostly unobstructed site, short "snapshot" experiments representing restricted solar exposures (sunlit narrow urban canyon, partial tree shade, and full shade), and long-sleeve shirt tests. The extended exposure and shirt measurements were conducted on the roof of the Novus parking garage in Tempe, Arizona (33.422°N, 111.928°W, 373 m elevation, see **Fig.2a**) on four cloudless days (September 11$^{th}$, 13$^{th}$, and 18$^{th}$ of 2024 with "nude" manikin and on October 11$^{th}$ of 2024 with manikin dressed in various long-sleeve shirts). The "snapshot" experiments were conducted for about 30 minutes each on October 16$^{th}$ and 17$^{th}$ of 2024 on the ground floor of the Walton Center for Planetary Health (33.421°N, 111.926°W, about 343 m elevation), with full shade provided by the building in the courtyard, the partial shade provided by a Palo Verde tree, and narrow urban canyon being formed by a gap in the building (see **Fig.2b**). Throughout these experiments, the air temperature varied from 30°C to 39°C and humidity from 10 to 22% (see **Supplemental Material—SM**).

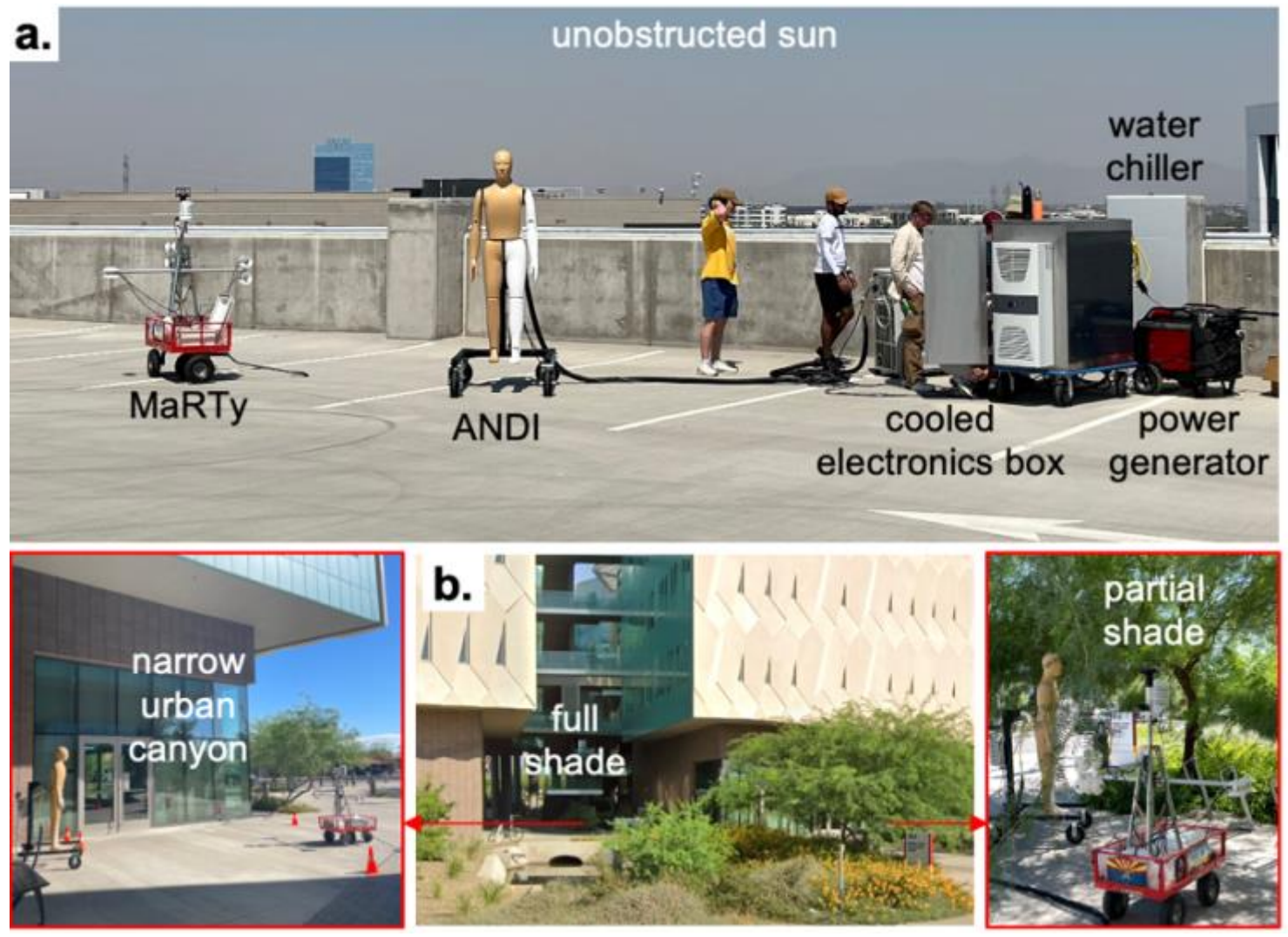

**Fig.2** The sites of the ANDI and MaRTy **a.** extended unobstructed sun exposure and **b.** snapshot experiments.

### *2.3 Two-color outdoor ANDI measurements and calculations*

The measurements were conducted using 35 zone outdoor thermal manikin ANDI system from Thermetrics LLC., which has dimensions that correspond to an average western male (see **Fig.1**). The system specifics and its outdoor operation are described in detail by Joshi et al. [43,111]. The left side foot, calf, thigh, pelvis, hand, forearm, upper arm, and underarm zones were coated with three thin layers of water-soluble water white paint (Premium Tempera, White Pint (72497), Blick #00011-1006). The calf, pelvis, forearm, and upper arm regions were divided into front and back zones (see **Fig.1f**), yielding eleven twin white-tan zone pairs. The coating was sufficiently thin not to impact the manikin temperature measurements, which we demonstrated using total heat flux measurements within a wind enclosure in a climate-controlled chamber [108] (see the **SM**). We measured the white paint's shortwave and longwave spectral properties using the procedure described by Rykaczewski et al. [63]. Integrating the spectral properties, presented in the **SM**, with emissive power distribution for a blackbody at 5800 K and 300 K yielded shortwave and longwave absorptivities of the white coating (referred to with subscript W) of $\alpha_{SW} = 0.15$ and $\alpha_{LW} = 0.98$, respectively. The manikin's original tan coating (referred to with subscript T) of the manikin has $\alpha_{ST} = 0.67$ and $\alpha_{LT} = 0.99$ [43]. The shortwave absorptivities of the two coatings were sufficiently different to cause substantial differences in measured heat fluxes. Their longwave absorptivities were close enough to be assumed equal, enabling the following calculations.

When the manikin is oriented towards the sun, the shortwave ($S_i$) and longwave ($L_i$) irradiation fluxes on the zones that have a symmetric "twin" across the sagittal plane can be determined from net heat flux measurements on the white ($q''_{i-W}$) and tan ($q''_{i-T}$) sides. During the outdoor experiments, we manually adjusted the manikin position every 5 to 10 minutes so that it faced the sun, as visually determined by the symmetry of the manikin's shadow (see **Fig.3a**). The convective heat transfer was eliminated by setting the exterior shell temperature for each zone ($T_{i-shell}$) equal to that of air, which is accomplished with manual input of the air measurement value into the ThermDAC software that then achieves this set point by balancing water cooling (the external water chiller was typically set to 5 to 10°C below air temperature) and internal heater power (see **Fig.1b**) [43]. To facilitate air-shell temperature matching, long exposure experiments were conducted in the afternoon when the air temperature was very stable (see average ANDI shell and air temperature variations in the SM) [43]. With convection eliminated, the energy rate balances on the two corresponding white-tan zones "$i$" are:

$$q''_{i-T} = \alpha_{ST} S_i + \alpha_{LT} L_i - \alpha_{LT} \sigma T^4_{i-shell} \tag{1}$$

$$q''_{i-W} = \alpha_{SW} S_i + \alpha_{LW} L_i - \alpha_{LW} \sigma T^4_{i-shell} \tag{2}$$

Since $\alpha_{LW} = \alpha_{LT}$, subtracting Equations 1 and 2 and rearranging, we obtain:

$$S_i = \left(q''_{i-T} - q''_{i-w}\right)/(\alpha_{ST} - \alpha_{SW}) \tag{3}$$

$$L_i = \left(q''_{i-T} - \alpha_{ST} \frac{(q''_{i-T} - q''_{i-w})}{(\alpha_{ST} - \alpha_{SW})} + \alpha_{LT} \sigma T^4_{i-shell}\right)/\alpha_{LT} \tag{4}$$

Where $\sigma$ is the Stefan-Boltzmann constant. It is important to highlight that this analysis assumes that the irradiation on the corresponding white and tan zones is equal. This assumption implicitly neglects that the shortwave irradiation reflected from the white part is higher than from the tan part and might contribute to an increase in shortwave irradiation on tan parts. For example, a fraction of the reflected shortwave radiation from the inner white thigh is absorbed by the inner tan thigh, increasing the measured heat flux. In the following sections, we will numerically investigate the magnitude of these self-reflections on the irradiation measurements.

We also developed an alternative approach to these measurements that was based on periodically changing the shortwave absorption of the entire manikin surface by briefly dressing ANDI with a custom full-body white onesie made out of white stretch fabric (Scuba double knit with wicking capabilities, L08-23-C3, Moodfabrics #312414, see **Fig.3b**). The underlying theoretical premise was the same as above, only with using heat fluxes measured sequentially for each zones with and without the white onesie, rather than those measured simultaneously for symmetric white-tan zones. Convection on the exterior surface of the onesie was eliminated by subcooling ANDI's surface below the air temperature. The temperature matching was accomplished with a custom code extension of the manikin control software considering the local fabric resistance measured within a climatic chamber. However, in trials conducted with an artificial light source (see **Fig.3b**), we found that under intense irradiation, the time required to dress (or undress) the manikin and switch between the two shell temperatures (for the exterior surface being equal to that of air with and without the onesie) was too long for outdoor measurements where solar angles change relatively quickly (see further details in the **SM Fig.S3**). Consequently, we settled on the symmetric white-tan zone variation of the two-color method.

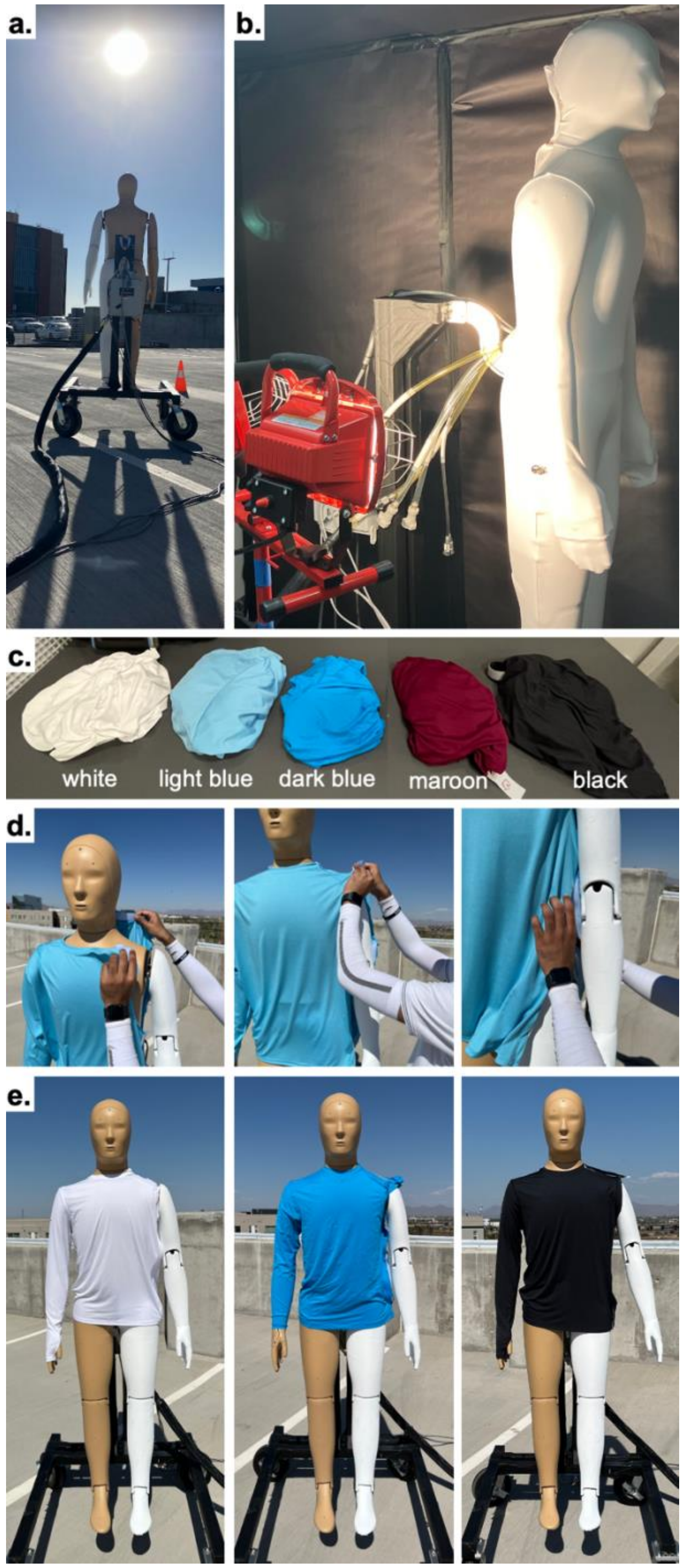

**Fig.3 a.** Image illustrating shadow-based alignment of the manikin with respect to the sun and **b.** the alternative shortwave and longwave irradiation method based on sequential measurements with and without the illustrated white onesie; images of the **c.** five tested long-sleeve shirts, **d.** the rapid dressing of the manikin using the modified shirts, and **e.** example images of ANDI dressed in white, dark blue, and black shirt.

### *2.4 Measurement of irradiation attenuation by long-sleeve shirts with varied colors*

To test how textiles of various colors attenuate the total irradiation on the manikin skin, we purchased and modified five long-sleeve shirts to allow for rapid changing and testing in outdoor conditions. In particular, we purchased a white (DSG, RN#104141, 100% Polyester, large, $\alpha_S = 0.05$, $\alpha_L = 0.97$), light blue (bloq-UV, 87% Nylon-13% Lycra Spandex, large, RN#135169, $\alpha_S = 0.29$, $\alpha_L = 0.98$), dark blue (bloq-UV, 87% Nylon-13% Lycra Spandex, large, RN#135169, $\alpha_S = 0.48$, $\alpha_L = 0.98$), maroon (DSG, RN#104141, 100% Polyester, large, $\alpha_S = 0.46$, $\alpha_L = 0.98$), and black (DSG, RN#104141, 100% Polyester, large, $\alpha_S = 0.53$, $\alpha_L = 0.98$) shirts with comparable thickness (see **Fig.3c**). We measured and calculated the shortwave and longwave absorptivities of the textiles using the same approach as for the white coating (see the **SM Fig.S4**). We had to rapidly dress and undress the manikin to ensure that the shirts were exposed to equivalent irradiation in outdoor conditions. To do so, we cut off the left sleeves and cut the remaining shirt on the right side from the neck down via the underarm region to the pelvis area. This approach allowed for "slipping" the shirt onto the manikin from the side via the right arm. To secure the shirt on the manikin, we attached matching parts of velcro tape (Hook&Loop Adhesive tape, 1" by 30 ft, Trazon) on the inner side of the cuts (see **Fig.3d**). We conducted three sets of shirt tests during afternoon of October 11$^{th}$ of 2024 on the unobstructed sun site (see **Fig.2a**), each involving a baseline nude measurement, followed by white-light blue-dark blue-maroon-black shirt testing. The shirt testing order was reversed on the second set to detect any potential "hysteresis" effect. While the "nude" experiments did quantify the shortwave and longwave radiation, the "dressed" parts of the manikin only quantified the net radiative heat flux (since shell temperature was set to the air temperature, convection did not contribute to the heating of the manikin). After changing each shirt, we monitored the measured heat fluxes across all the relevant zones until they stabilized (typically about two minutes) and recorded two minutes of steady-state data. We note that to facilitate interpretation, in these experiments, MaRTy was realigned along with ANDI to face the sun (i.e., only one vertical detector faced the sun directly).

### *2.5 MaRTy measurements and 6-directional flux transformation into directional and diffuse sources*

MaRTy measures shortwave and longwave irradiation fluxes in cardinal and vertical directions [64] that, for very anisotropic cases such as outdoor solar radiation, must be transformed into the primary directional and diffuse sources to avoid artificial over-shading and over-exposure on complex shapes such as the human body [88]. We will employ measurements from the unobstructed sun site conducted on the afternoon of September 18th, 2024, to illustrate the transformation calculations (see **Fig.4a**).

Following Holmer [112], we assume that the isotropic diffuse source of shortwave radiation, which includes both the uniformly diffuse atmospheric (does not include circumsolar and horizon zones as in the Perez model [113]) and diffusely reflected radiation contributions, is at any given time the minimum of the four vertical sensors:

$$S_{diff} = \text{Min}\ (S_{east}, S_{west}, S_{north}, S_{south}) \tag{5}$$

In the afternoon case, the shortwave diffuse signal corresponds to the north-facing sensor ($S_{north}$). Confirming that this is indeed the "background" radiation level, this signal is also nearly equal to that from the east-facing sensor ($S_{east}$), which is also not exposed to the sun in the afternoon (see schematic of the top-down view in **Fig.4c**). For highly absorbing or fully diffusively reflecting surfaces, the $S_{diff}$ is also equal to the signal measured by the downward-facing detector ($S_{down}$) [62]. In contrast, the $S_{down}$ in **Fig.4a** is substantially higher than $S_{diff}$ (i.e., $S_{north}$), which implies that the light concrete pavement reflects more shortwave radiation in the direction facing the sun. For simplicity, we assume that this additional radiation originates from a specular component of the reflected direct shortwave radiation ($S_{dir}$ from the ground at angle $\theta$ from the surface normal, where $\theta$ is the solar zenith angle, see **Fig.4d**) and is characterized by a reflectivity $\rho$ (i.e., is equal to $\rho S_{dir}$). Therefore, in our model, the upward- and downward-facing shortwave sensor signals are $S_{up} = S_{dir} cos\theta + S_{diff}$ and $S_{down} = \rho S_{dir} cos\theta + S_{diff}$, respectively. Knowing the zenith angle and $S_{diff}$, we can calculate the direct shortwave radiation as:

$$S_{dir} = \left(S_{up} - S_{diff}\right)/cos(\theta) \tag{6}$$

While the value of the directional reflectivity can be determined using:

$$\rho = \frac{\rho S_{dir} cos\,(\theta)}{S_{dir} cos\,(\theta)} = \frac{S_D - S_{diff}}{S_U - S_{diff}} \tag{7}$$

**Fig.4e** shows $S_{diff}$, $S_{dir}$, and $\rho S_{dir}$ calculated using **Eq. 5-7** for the September 18th afternoon experiments, while the inset in the figure shows the $\rho$ calculated for experiments. Besides the last ~30 minutes before sunset, the $\rho$ values in all cases are nearly constant at about 0.15 to 0.17 (as are for repeated experiments on September 11th and 13th; see the **SM Fig.S5**). The decrease in the value right before sunset might be physical or an artifact of very low shortwave radiation and shallow solar angle in reference to the two horizontal sensors. The latter aspect also mandates a modification of how $S_{dir}$ is calculated for zenith angles above 60°, as described next.

The calculation of $S_{dir}$ should be based on the sensor that receives the most direct radiation, which in the late afternoon is the signal from the west-facing vertical sensor ($S_{west}$). According to the direct, reflected, diffuse shortwave model, $S_{west} = \left((1+\rho)S_{dir} sin\theta + S_{diff}\right) sin\phi^*$, where local azimuth angle, $\phi^* = \phi + \Delta - 180°$, accounts for potential MaRTy misalignment from cardinal direction by angle $\Delta$ that can be quantified from the cross-over of the east and west facing sensors signals in terms of the azimuth angle [112] (see the **SM Fig.S6**). Similarly, $S_{south} = \left((1+\rho)S_{dir} sin\theta + S_{diff}\right) cos\phi^*$. Confirming the need to use the alternative model for $S_{dir}$ for high zenith angles, the model using inputs from **Eq.5-7** starts to underpredict $S_{west}$ as compared to measured values for $\theta$ over 60°, with the highest discrepancy reaching around 150 $W \cdot m^{-2}$ near sunset (see the **SM Fig.S7**). Rearranging the expression for $S_{west}$, we obtain the formula for $S_{dir-2}$ for late afternoon (or early morning if $S_{east}$ is used):

$$S_{dir-2} = \frac{\left(\frac{S_{west}}{sin\phi^*} - S_{diff}\right)}{(1+\rho) sin\theta} \tag{8}$$

The plot in **Fig.4e** compares the direct solar irradiation for September 18th calculated using **Eq.6** and **Eq.8**, with the latter including either varying (as shown in the inset) or constant (0.16) value for $\rho$. These transformed measurements are also compared against the Haurwitz global solar radiation model (radiation flux on a horizontal surface) [88,114]:

$$S_{g-theory} = 1367 k_t cos\theta e^{-\frac{0.057}{cos\theta}} \quad (9)$$

Where $k_t$ is the clearness index that for our location we assume to be 0.8 [88,114]. The direct and diffuse components of the solar irradiation can be obtained from **Eq.9** as $S_{diff-theory} = k_d S_{g-theory}$ and $S_{dir-theory} = (1 - k_d) S_{g-theory}/cos\theta$ [114], where $k_d$ is the diffuse fraction that for our location we assume to be 0.2 [88,114]. In terms of the global irradiation, the $S_{g-theory}$ agrees well with "measured" $S_g = S_{dir} cos\theta + S_{diff}$ obtain using inputs from **Eq.5-6**. However, when decomposed into the direct and diffuse components, the measured values transformed using **Eq.5-6** are lower than the theoretical past about 16:00 (i.e., $\theta$ of about 60°). At this time, the $S_{dir}$ calculated using **Eq.6** and $S_{dir-2}$ using **Eq.8** overlap. Later on, their curves diverge, with $S_{dir-2}$ being 120 to 150 $W \cdot m^{-2}$ higher near the sunset. Using a constant or varying value of $\rho$ only makes a difference near sunset when the variable $\rho$ decreases (see inset in **Fig.4e**). It is worthwhile pointing out even that $S_{dir-2}$ with varying $\rho$ in the late afternoon is closer but still lower than values predicted by the Haurwitz model. This feature might stem from using a single value for the clearness index, which might be higher in the late afternoon when sun rays are more impacted by air pollution above the urban areas. In summary, we will calculate the direct shortwave radiation from the 6-directional MaRTy fluxes using **Eq.6** for $\theta$ below 60°, while for higher values, we will apply **Eq.8**. Lastly, in contrast to the varying shortwave fluxes, the longwave fluxes are constant across the afternoon with all vertical sensors at about 475 $W \cdot m^{-2}$, the upward-facing sensor at 375 $W \cdot m^{-2}$, and the downward-facing sensor at 575 $W \cdot m^{-2}$ (see **SM Fig.S7**).

We also used the 6-directional fluxes to calculate the whole-body absorbed radiation ($R_a$) by weighting the shortwave ($S_j$) and longwave ($L_j$) irradiation fluxes measured using MaRTy with standard absorptivities ($\alpha_S$ of 0.7 for shortwave and $\alpha_L$ of 0.97 for longwave radiation [56]) and geometrical weights ($W_j$) corresponding to a single-cylinder human [62,64] as follows:

$$R_a = \sum_{j=1}^{6} W_j \left(\alpha_s S_j + \alpha_L L_j\right) \quad (10)$$

The geometrical weight factor for the cardinal directions is 0.22 and 0.06 for the upwards and downwards irradiation. We calculated the Mean Radiant Temperature (MRT in degrees Celsius) from $R_a$ as:

$$MRT = \sqrt[4]{\frac{R_a}{\alpha_L \sigma}} - 273.15 \quad (11)$$

Where $\sigma$ is the Stefan-Boltzmann constant equal to $5.67 \times 10^{-8}$ W·m$^{-2}K^{-4}$.

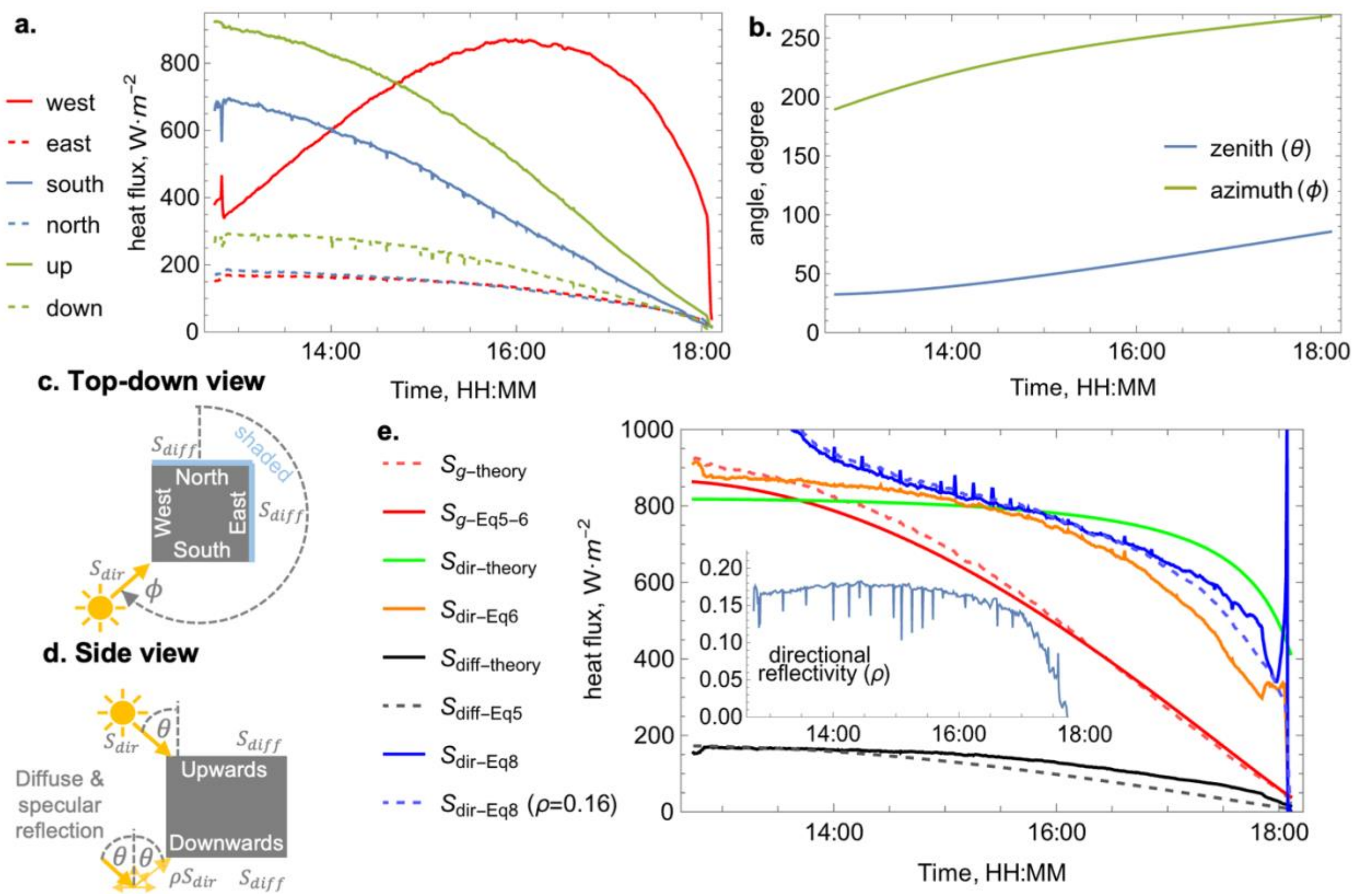

**Fig.4 a.** The shortwave fluxes measured using MaRTy during the afternoon on September 18th, 2024, **b.** the corresponding changes in solar zenith and azimuth angles, **c.** top-down, and **d.** side view of the detector orientation and direct, reflected, and diffuse shortwave radiation exposure indicated, and **e.** the variation in the global, direct, and diffuse shortwave irradiation calculated using **Eq.5-9** and data from a&b; inset shows variation in the directional reflectivity calculated using **Eq.7**.

### *2.6 Virtual ANDI twin radiation simulations*

Since we experimentally eliminate convection, we focus on simulating the anisotropic irradiation imposed by a combination of directional and diffuse radiation sources on an isothermal and opaque virtual manikin with absorptivity/reflectivity values matching the tan/white distribution on the physical manikin. As described in depth in our prior simulation work [88,91], we conducted all simulations using the Surface-to-Surface Radiation node in the Heat Transfer module of Comsol Multiphysics 6.1. The virtual ANDI twin corresponds directly to the physical manikin and its zones, with minor geometrical simplifications performed to facilitate the meshing process, as described by Viswanathan et al.[115]. We utilized "normal" triangular mesh on the surface of the manikin (30258 elements), as switching to "fine" (64966 elements) and "finer" (153204 elements) made under 0.1% difference in simulated irradiation per each of the zones illustrated in **Fig.1f** (see the final mesh in **Fig.5a**). As demonstrated experimentally below, the longwave irradiation across the manikin was nearly uniform across all the zones. Consequently we focused on simulating the shortwave exposures (the longwave emission was artificially eliminated by setting the manikin surface temperature to 0 K under the “diffuse surface” setting)

For each simulated shortwave irradiation case, we imported the $\theta$, $\rho$, $S_{dir}$ (or $S_{dir-2}$), and $S_{diff}$ calculated from MaRTy measurements with a 15-minute spacing using **Eq.5-8**. The manikin's surface was modeled as "Diffuse Surface" with emissivity corresponding to the shortwave absorptivity of the tan or white zones. The $S_{dir}$ (with a downward direction at an angle of $\theta$ from the normal to the horizontal plane) and $\rho S_{dir}$ (with an upward direction at an angle of $\theta$ from the normal to the horizontal plane) directional shortwave sources were simulated using the "External Radiation Source" with an "Infinite distance" position facing the manikin. To model the $S_{diff}$ source, we utilized the "Ambient" radiation under the "Diffuse Surface" sub-node with an equivalent ambient temperature" calculated as $\sqrt[4]{S_{diff}/\sigma}$ (in degrees Kelvin). We selected the "Ambient" radiation approach over the "Diffuse irradiance" option because the latter produces a uniform irradiation over the entire manikin surface, regardless of its topography. The "Ambient" radiation produces a more realistic diffuse environmental irradiation that accounts for self-shading effects (e.g., the reduced irradiation on the underarm region shaded by the arm—see **Fig.5b-d**).

In the numerical solution, we employed the Hemicube method with 64 resolution [91], with multiple cases simulated sequentially with the Matlab livelink code passing in the shortwave values and exporting the derived values for each zone into a .csv file. The zone-area averaged values of the external ($S_{dir}$), ambient ($S_{diff}$), and mutual (reflected from one zone onto the other) were obtained using "Derived Values" with "rad.Gext", "rad.Gamb", and "rad.Gm_gp" expressions (see example distributions on the manikin in **Fig.5b-d**). To quantify the impact of the mutual reflections on the measured heat flux, we compared total irradiation fluxes calculated as "rad.Gext+rad.Gamb" and "rad.Gext+rad.Gamb+rad.Gm_gp". In particular, to directly compare the simulations involving reflections to the measurements, we calculated the equivalent flux from simulations for the tan and white zones and combined them according to **Eq.3**. To compare the virtual ANDI simulations against the IRM method, we calculated the total shortwave irradiation flux across the manikin both with and without the mutual reflections. The longwave irradiation was assumed to be uniform across the manikin and equal to the average of the 6-directional MaRTy fluxes. The resulting $R_a$ was substituted into **Eq.11** to calculate the equivalent MRT.

### *2.7 Simple model for radiation attenuation provided by clothing*

The zonal shortwave and longwave fluxes, whether directionally measured using the two-color ANDI method or obtained through virtual manikin simulation using MaRTy fluxes, can be used torelate the spectral fabric properties to its cooling performance. In particular, the $S_i$ and $L_i$ irradiation fluxes can be related to the $q''_{shirt}$ and $\alpha_S$ and $\alpha_L$ of the fabrics (indicated here with "f" subscript) through the following rudimentary model (with temperatures expressed in Kelvin):

$$q''_i = \alpha_{L-f}\sigma T_{i-f}^4 - \alpha_{L-shell}\sigma T_{i-shell}^4 \quad (12)$$

Where the $T_{i-f}$ is the temperature of the fabric that can be estimated using heat rate balance on the textile:

$$\alpha_{S-f}S_i + \alpha_{L-f}L_i + \alpha_{L-shell}\sigma T_{i-shell}^4 = 2\alpha_{L-f}\sigma T_{i-f}^4 + 2h\left(T_{i-f} - T_{air}\right) \quad (13)$$

We note that in this simplistic model, we assume the shirt to be fully opaque, is cooled equally on both sides through convection with coefficient $h$, and has the same surface area as the underlying manikin shell zone. We selected the front upper arm zone because it has directly measured spectral fluxes, a relatively large and sun-exposed area (e.g., compared to the small shoulder zone), and has minimal disturbance due to the fabric mounting method.

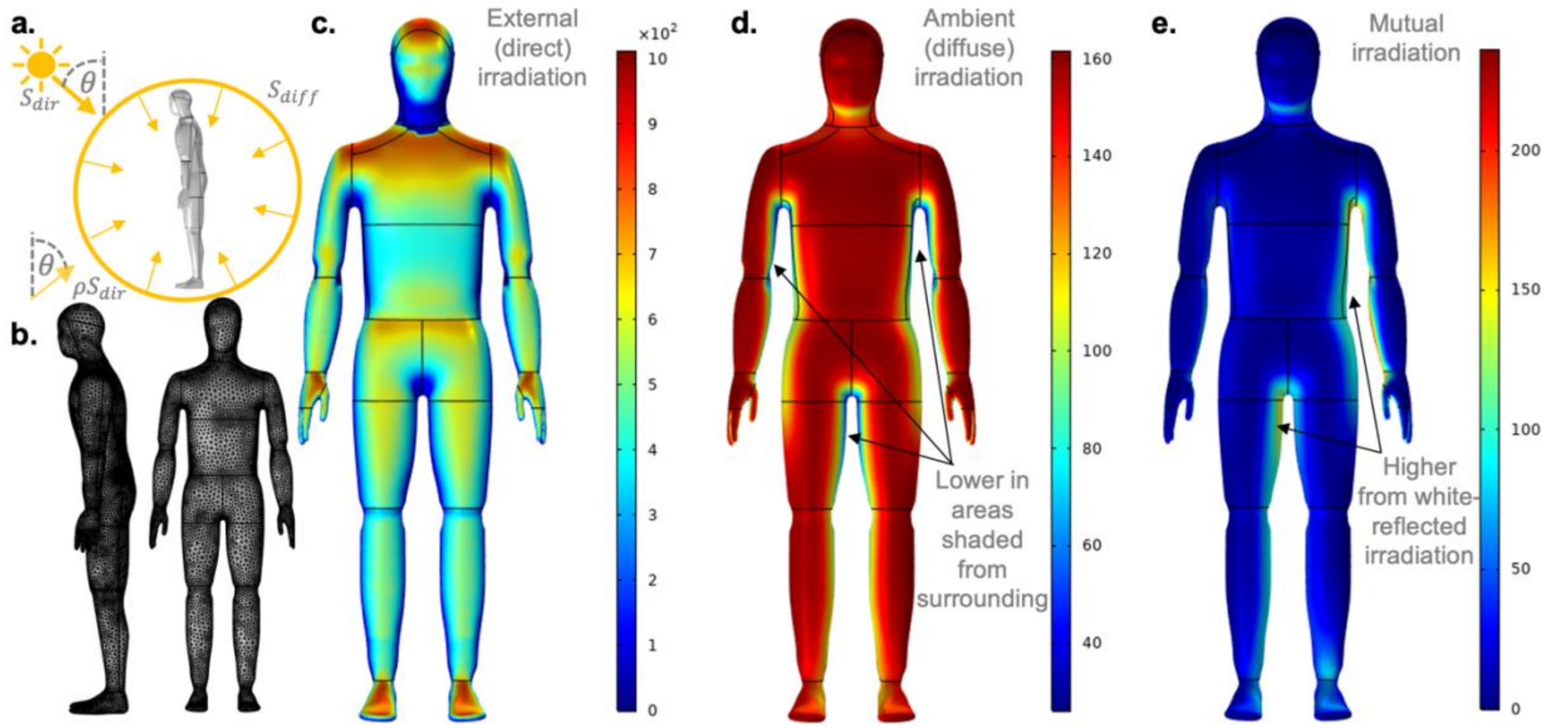

**Fig.5 a.** Illustration of the direct, reflected, and diffuse radiation sources the virtual manikin is exposed to, **b.** the virtual ANDI mesh; **c-e.** example simulated **c.** external ($S_{dir}$), **d.** ambient ($S_{diff}$), and **e.** mutual (reflected from one zone to another, with the majority excess coming from reflection from white zones) shortwave irradiation for $\theta$ =39.2°, $\rho =$ 0.14, $S_{dir}$ = 851.2 W·m$^{-2}$, $S_{diff}$ = 161.7 W·m$^{-2}$ (equivalent ambient temperature of 231.1 K); Units on all color legends are W·m$^{-2}$.

## 3. Results

### *3.1 Extended experiments in the unobstructed sun conditions*

While the longwave fluxes measured using MaRTy and ANDI were nearly constant throughout the three extended exposure afternoon measurements (see **Fig.6a**), the direct shortwave radiation decreased nearly threefold through the same time (see **Fig.6b**). In particular, the MaRTy longwave fluxes measured in the cardinal directions were nearly equal and only decreased from ~490 to 450 W·m$^{-2}$ during five hours, while the fluxes from the downward and upwards facing sensors were 100 W·m$^{-2}$ higher and lower, respectively. The longwave radiation fluxes measured on all of ANDI's twelve zones with white-tan pairs essentially overlapped the MaRTy measurements. They were within ±25 W·m$^{-2}$ of the average (see the fluxes measured on September 18th in **Fig.6a,** which are representative of the three days). Therefore, longwave irradiation can be considered uniform across ANDI's body, and we focus on comparing measurements and simulations of shortwave irradiation.

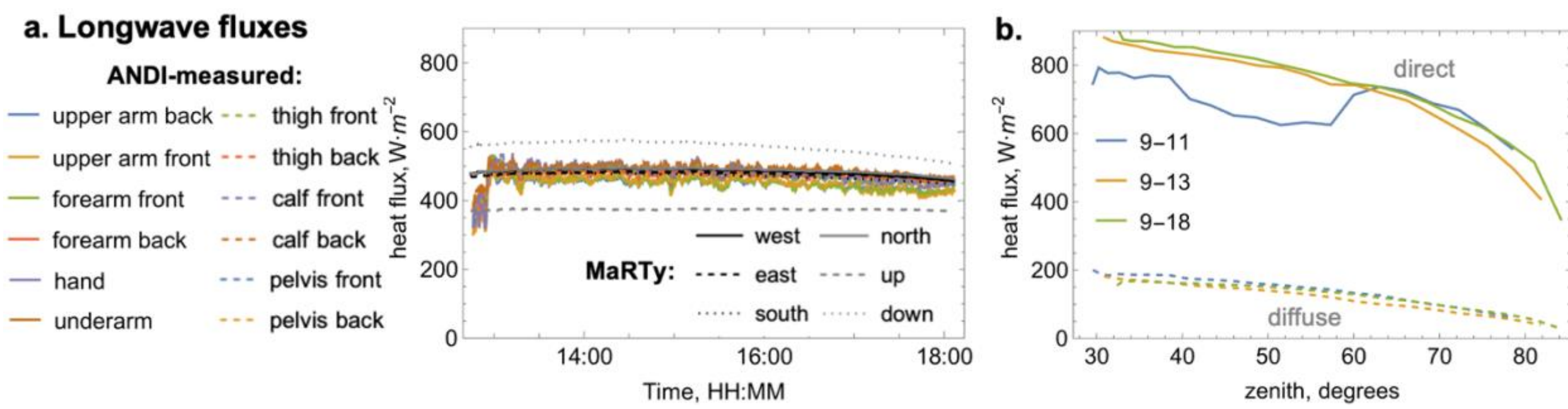

**Fig.6 a.** Comparison of longwave fluxes measured on ANDI's zones with white-tan pairs and using MaRTy on September 18th, 2024 on the unobstructed roof site, and **b.** comparison of direct and diffuse shortwave radiation ($S_{dir}$ and $S_{diff}$) calculated using **Eq.5-8** using MaRTy data for September 11th, 13th, and 18th, 2024 with a 15-minute frequency used for simulations on the unobstructed roof site (as a function of the solar zenith angle to eliminate daily shifts due to changing solar angles).

The plot in **Fig.6b** shows that the direct and diffuse shortwave radiation calculated using **Eq.5-8** from MaRTy measurements nearly overlaps on the three afternoons, except the 150 to 200 W·m$^{-2}$ dip in the early afternoon on September 11th. This decrease in the direct shortwave radiation corresponded to unusually high air pollution with smoke and haze from the Log Angeles National Forest fire passing over Tempe. These shortwave flux values and directional reflectivity (see **Fig.4e** and **SM Fig.S5**) were input into the simulation to yield irradiation values across the "virtual ANDI".

**Fig.7** shows the measured and simulated shortwave irradiation fluxes across ANDI's twelve zones with white-tan pairs agree well in most cases. The results for the two clear days on September 13th and 18th match closely, with a significant (>100 $W \cdot m^{-2}$) disagreement between simulations and measurements observed only for the front part of the pelvis and the hand. The September 11th "natural experiment" results with pollution-attenuated directed shortwave radiation demonstrate that the simulations can predict complex temporal changes in the measured irradiation. All the zones on the back of the manikin, shaded from the direct irradiation component, closely match the diffuse shortwave value. Only the underarm region is exposed to lower fluxes than the back zones, at most 100 to 150 $W \cdot m^{-2}$. The front zones are exposed to relatively constant shortwave fluxes from 13:00h to about 17:00h with values between 300 and 500 $W \cdot m^{-2}$. We note that the sudden drop in the fluxes measured on the back of the pelvis and thigh around 16:00 h was due to manual adjustment of the water chiller temperature to account for the large heat load. Demonstrating a minor impact of mutual exposure, the difference between simulations with and without reflections is typically lower than 10 $W \cdot m^{-2}$, with the discrepancy reaching at worst 30 to 40 $W \cdot m^{-2}$ (or under 10%) for the front parts of the calf and thigh. Next, we discuss measurements and simulations performed in various sun-restricted cases.

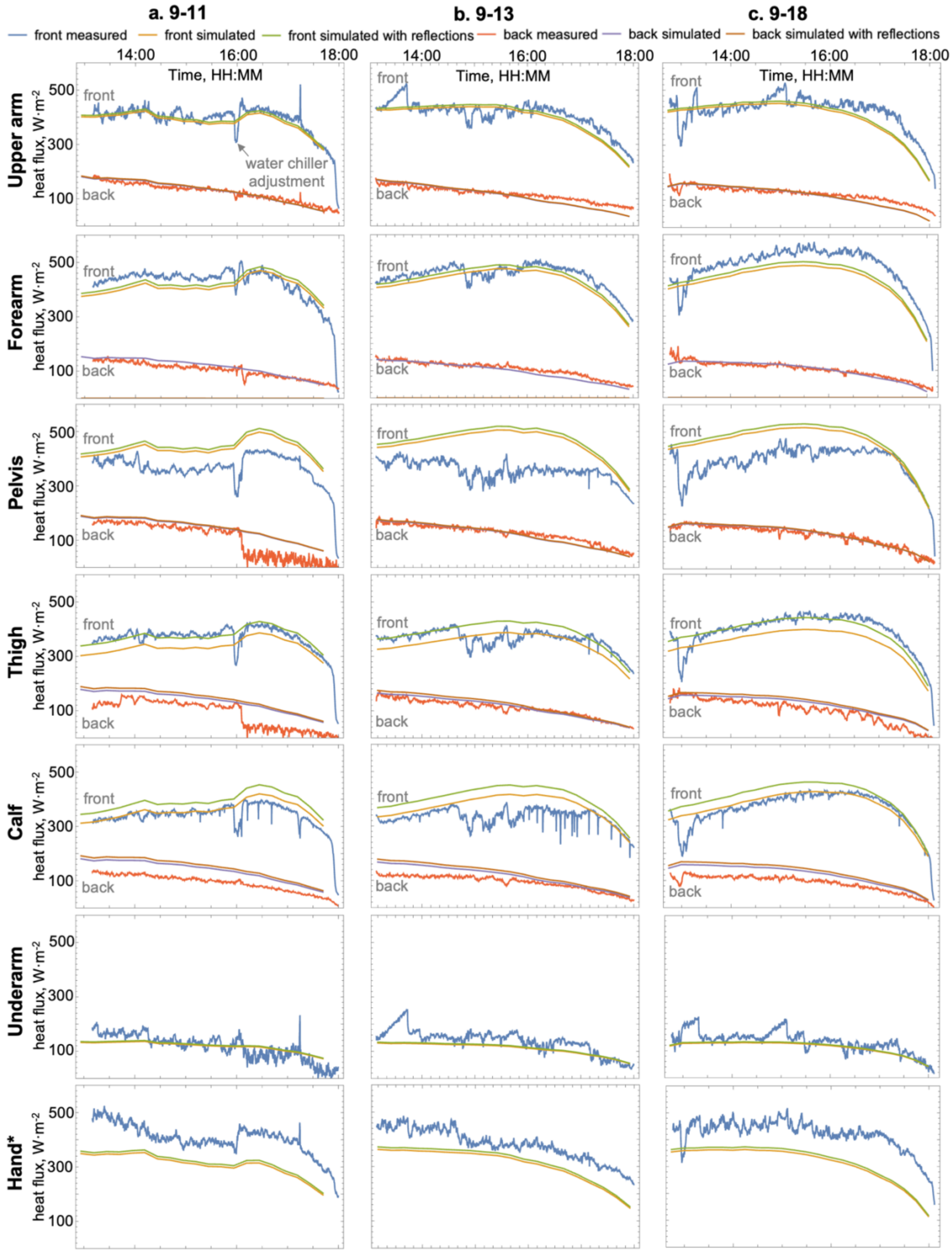
a. 9-11
b. 9-13
c. 9-18
front measured
front simulated
front simulated with reflections
back measured
back simulated
back simulated with reflections
14:00
16:00
18:00
Time, HH:MM
Upper arm
Forearm
Pelvis
Thigh
Calf
Underarm
Hand*
heat flux, W·m-2
500
300
100
front
back
water chiller adjustment

**Fig.7** Comparison of ANDI-measured and simulated shortwave irradiation across the manikin's zones during experiments conducted on September **a.**11$^{th}$, **b.**13$^{th}$, and **c.**18$^{th}$, 2024 on the unobstructed roof site. The simulations include simulated shortwave irradiation with and without considering mutual reflections between the manikin zones. *the hand zone does not include fingers.

### *3.2 Snapshot experiments in restricted solar exposures*

The measurements conducted in full shade provided by a building represent the opposite extreme of the unobstructed sun-site conditions and confirm the capabilities of the two-color ANDI method. In particular, all MaRTy and nearly all ANDI measured shortwave fluxes were below 30 W·m$^{-2}$, while the longwave fluxes were around 490 W·m$^{-2}$. Since all the fluxes remained steady throughout the 30-minute measurement period (see **SM Fig.S8**), we present the time-averaged values in the bar plot in **Fig.8a**. From this plot, it is evident that only the shortwave flux measured on the hand is an outlier that exceeds any MaRTy measured values. Having demonstrated the viability of the two-color ANDI method in extremely sunny and fully shaded conditions, we will discuss the results for the two in-between scenarios.

Within the narrow urban canyon, the simulated shortwave zonal fluxes overestimate the ANDI-measured values across all zones except the hand. Specifically, the bar plot in **Fig.8b** summarizing the 5-minute measurement period, during which the values were steady (see **SM Fig.S9**), shows that besides being higher for the front of the pelvis, which is consistent with the unobstructed site results, the simulations also substantially exceed measurements for the front of the upper arm (by about 100 W·m$^{-2}$) and calf (by 50 to 75 W·m$^{-2}$) as well as for most of the back zones (by 20 to 50 W·m$^{-2}$). Consistent with the unobstructed site results, the hand shortwave measurement exceeds the simulations by about 60 W·m$^{-2}$. As in the case of the full shade, the longwave fluxes measured by both MaRTy and ANDI in the narrow urban canyon and in partial tree shade were all nearly identical at around 490 W·m$^{-2}$ (see **SM Fig.S9** and **Fig.S10**).

In contrast to the steady values of the shortwave fluxes in the fully shaded and the narrow urban canyon, the shortwave fluxes from the sun-facing (north- and upward-facing) MaRTy sensors varied highly during the 30-minute measurement period in partial shade provided by the Palo Verde tree (see **Fig.8c**). The variation in these fluxes translated into highly varying direct shortwave radiation, and therefore also oscillating simulation results. However, the simulations underpredicted the shortwave measurements for all body parts by 25 to 100 $W \cdot m^{-2}$ (see upper arm example in **Fig.8d** and rest of the zones in **SM Fig.S11**). Next, we discuss the attenuation of the irradiation provided by dressing the manikin in long-sleeve shirts.

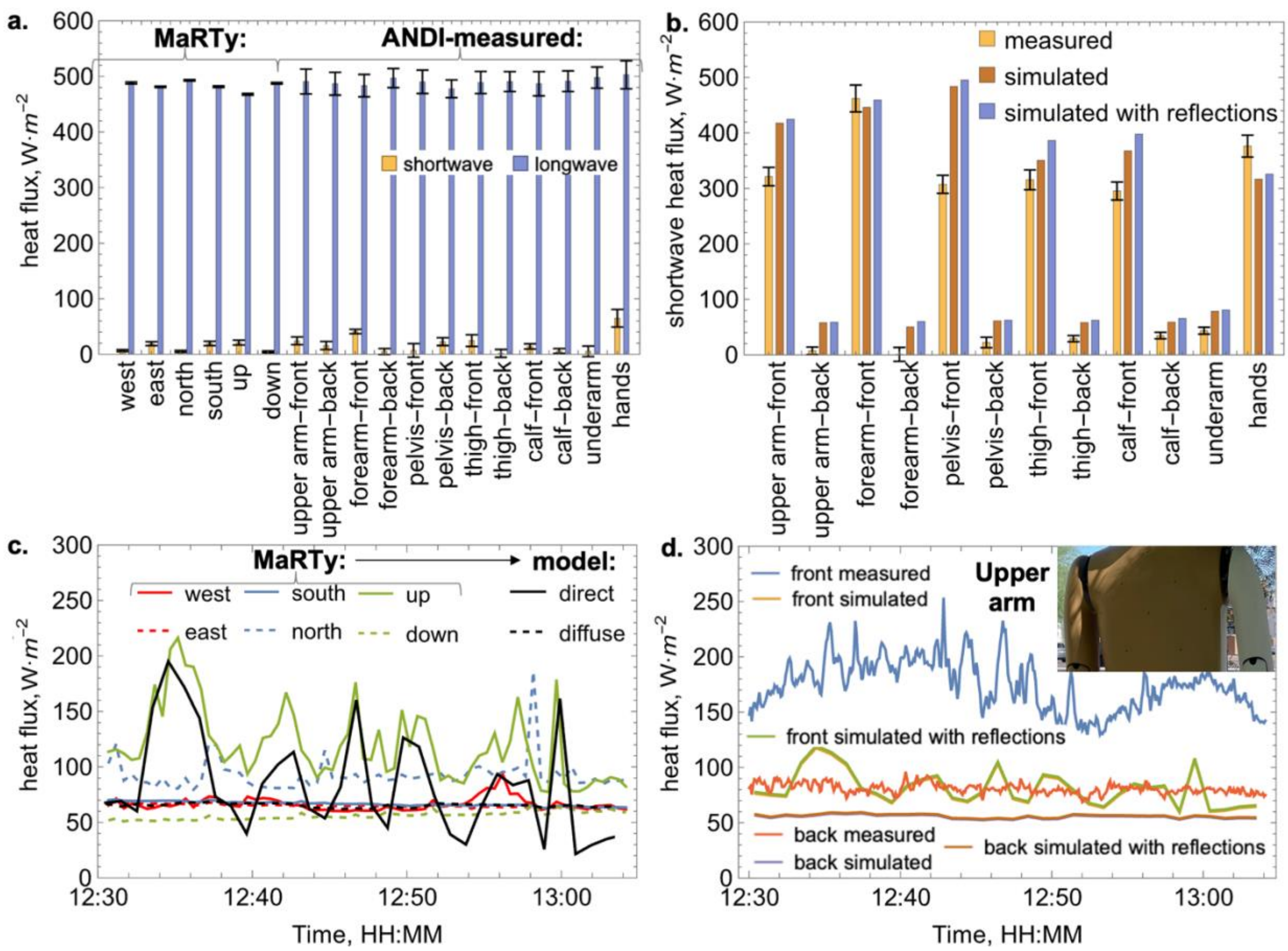

**Fig.8 a.** MaRTy- and ANDI-measured shortwave and longwave fluxes in full shade, **b.** ANDI-measured and simulated shortwave fluxes in the narrow urban canyon (error bars correspond to 90% confidence interval), **c.** the MaRTy measured shortwave fluxes along with corresponding direct and diffuse fluxes calculated using **Eq.5-8** in partial shade provided by Palo Verde Tree, **d.** comparison of ANDI-measured and simulated shortwave fluxes on the upper arm (example picture shown in inset) in the partial shade.

### 3.3 The attenuation of irradiation provided by long-sleeve shirts of varying color

The primary challenge in conducting experiments comparing the performance of multiple textiles in outdoor settings is ensuring that testing conditions remain constant. With each set of tests (baseline and five shirts) taking 45 to 60 minutes, the solar zenith angle changed by about 5° between the baseline measurements (i.e., nude manikin measurements before and after each shirt set—see **SM Fig.S12a**). The corresponding change in net radiative heat flux (i.e., $q''$) measured during the four baseline repetitions dependent on the zone but did not exceed 80 W·m$^{-2}$. In particular, the $q''$ remained constant across the four repetitions for all the back zones (upper arm, forearm, upper back, lower back), increased gradually for the front of the upper arm, forearm, underarm and stomach, while decreasing only for the shoulder (see **SM Fig.S12b**). The decomposition into spectral bands demonstrates that, as in all prior experiments, the longwave irradiation remained unchanged and equal across all the zones; therefore, the change in $q''$ was due to shortwave variation (see **SM Fig.S12c-d**). Out of the four baseline measurements, the changes in the $q''$ between the 3$^{rd}$ and 4$^{th}$ repetitions was minimal across all the zones (see **SM Fig.S12c-d**). Accordingly, we present below the fluxes measured with the five shirts in between these baselines but note that the results for the other two repetitions,and on average across all repetitions, were very comparable (see **SM Fig.S12e**).

Wearing any of the shirts reduced the net radiative heat flux compared to the "bare" tan shell across all zones, with the lighter color textiles providing the largest reduction. For example, the bar plot in **Fig.9a** shows that wearing a white and black shirt for the stomach zone reduced baseline 460 W·m$^{-2}$ to 203 W·m$^{-2}$ and 320 W·m$^{-2}$, respectively. To facilitate the comparison across the different zones, we scaled the fluxes measured with different shirts by the baseline bare tan shell flux (i.e., present the "scaled flux"). The results in the bar plot in **Fig.9b** demonstrate that wearing the white shirt reduced the scaled flux to 0.3 to 0.5 across all zones except the underarm (this mostly shaded zone had a scaled flux of 0.75). The light blue shirt also greatly reduced the scaled flux with values of 0.4 to 0.6. In contrast, the scaled fluxes measured while manikin was wearing the three darker shirts (dark blue, maroon, and black) were comparable to each other and often 2 to 3 times higher than those measured with the white shirt. For example, the white shirt on the chest zone reduced scaled flux to 0.5, while the darker shirts reduced it to only about 0.9. The lowest scaled fluxes measured with the dark shirts were around 0.6 to 0.7 for the front of the forearm, the shoulder, the stomach, and the lower back. In absolute terms, when the fluxes for each zone are multiplied by their area and summed up across the upper body, the 159 W absorbed by the tan shell are reduced to 73.3 W, 86.7 W, 128 W, 117 W, and 127 W when wearing white, light blue, dark blue, maroon, and black long-sleeve shirts.

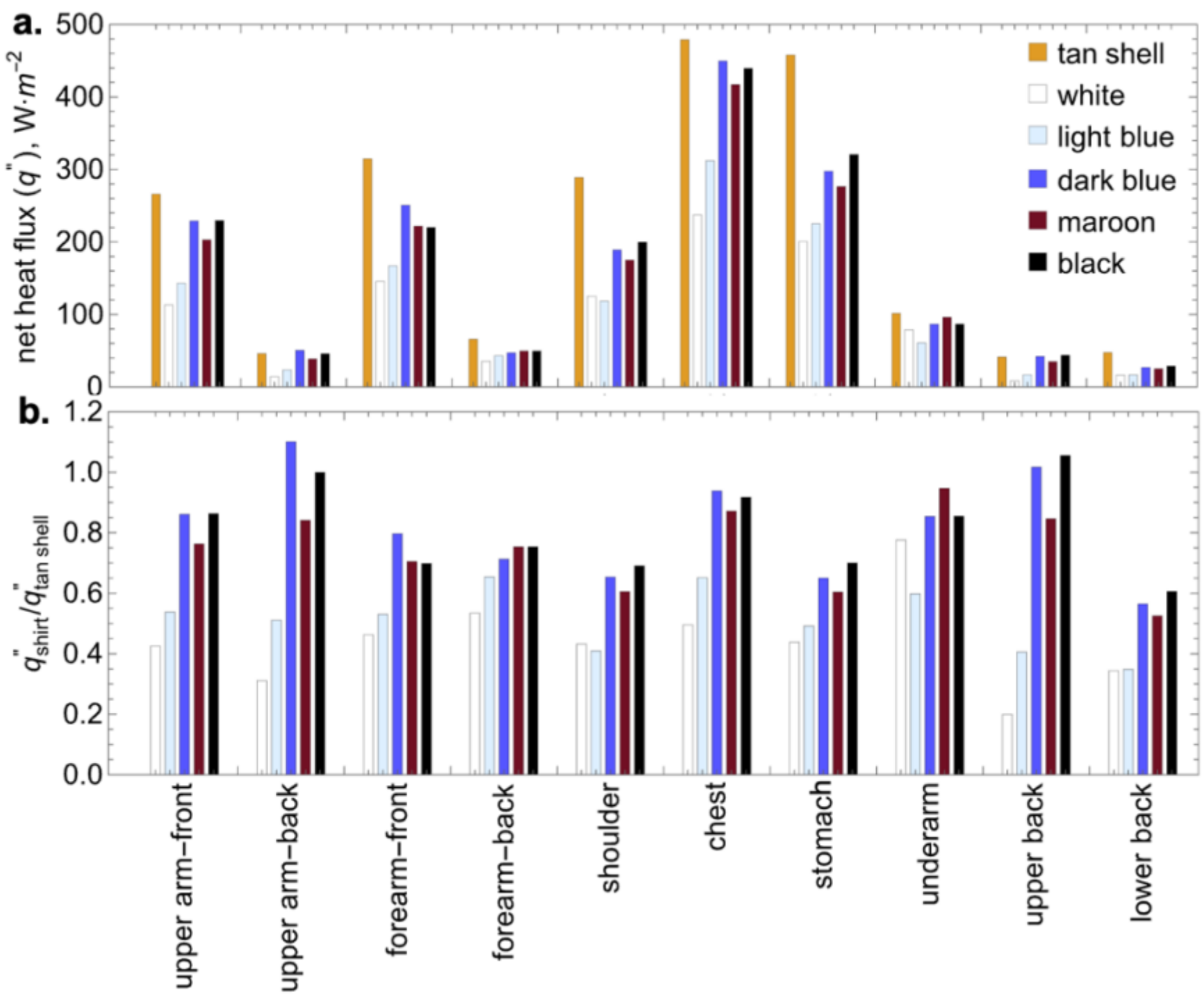

**Fig.9** The radiative heat flux across the upper body zones with and without the five long-sleeve shirts measured using ANDI (3rd repetition, see rest in **SM Fig.S12**) in **a.** flux and **b.** scaled flux ($q^{"}_{shirt}/q^{"}_{tan\,shell}$) terms.

### *3.4 Modeling radiation attenuation provided by clothing*

By substituting the conditions for the 3rd shirt testing repetition for the front upper arm zone ($S_i = 476$ W·m$^{-2}$, $L_i = 493$W·m$^{-2}$, $T_{air} = T_{shell} = 38$°C, and a $h = 10$ W·m$^{-2}$°C$^{-1}$ that is representative of the measured wind speed of 1.3±0.8 m·s$^{-1}$ [43]), we obtain the model values for the front upper arm presented in **Fig.10**. Despite the rudimentary nature of the model, it predicts $q^{"}_{shirt}$ values within 15% of the ANDI-measured net radiative fluxes.

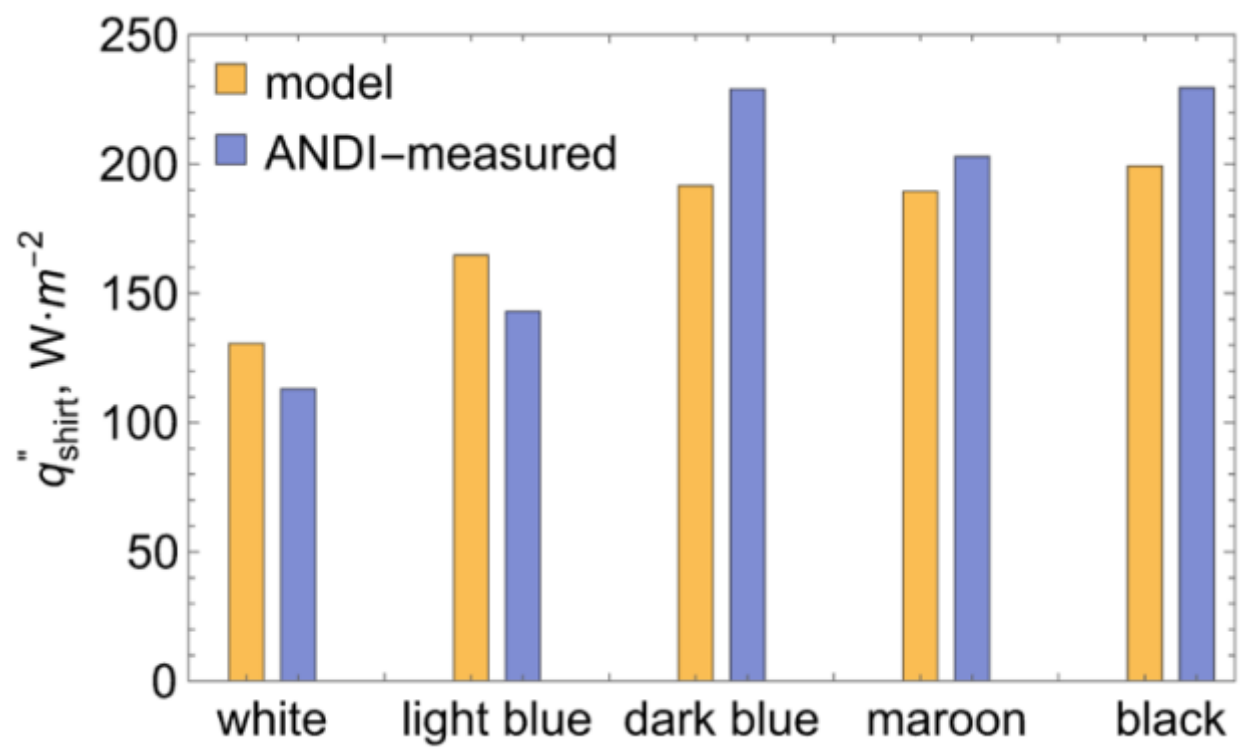

**Fig.10** The net radiative heat flux for the upper arm measured using ANDI and predicted using the model in **Eq.12-13** for manikin covered by the five long-sleeve shirts with indicated colors.

## 4. Discussion

### *4.1 The unobstructed sun conditions*

Our results demonstrate that the longwave irradiation can be considered uniform and equal to the 6-directional average across the manikin (both in full sun and full shade). In contrast, the shortwave irradiation measured using the two-color manikin method can be matched in most conditions using simulations, with and without considering mutual reflections, for all zones except the pelvis and hands. Using simulations, we demonstrated that mutual reflections have a minor contribution to fluxes measured using the two-color manikin method. The highest impact of the mutual reflections reached about 10% (30 to 40 $W \cdot m^{-2}$) for the front parts of the calf and thigh, which stems from the proximity and high view factors between the inner parts of the two legs. We observed the largest theory-to-experimental differences for the pelvis and hand zones. The simulations likely overestimate the shortwave irradiation on the pelvis because, to allow for actuation in mimicked walking motion, the top of the physical zone curves under and is partially shaded by the stomach zone (the small casted shadow is even visible in **Fig.1a**). This complex geometrical feature is not captured in the simplified virtual manikin, explaining why the simulations provide higher values for the pelvis than the measurements. Similarly, the simulated underpredictions of the shortwave irradiation measured by the hand zone have physical roots.

Only the palm of the physical ANDI is internally water-cooled and has integrated temperature and heat flux sensors. Accordingly, we initially only simulated palm-hands without the fingers (see **SM Fig.S13a**). Considering our results in **Fig.7** and **Fig.8**, we conducted additional simulations with the whole hand, showing a small reduction in the simulated palm heat flux (see **SM Fig.S13b**). In reality, the fingers are not entirely "passive" but also heat up and can contribute to the heat rate measured by the cooled palm region via heat conduction within the shell. To estimate such contribution, we calculated the shortwave flux that would be measured if palm sensors captured the entire irradiation from the fingers. In that case, rather than underestimating the palm measurement by about 100 $W \cdot m^{-2}$, the simulations would exceed the measurements by about 200 $W \cdot m^{-2}$. Naturally, a fraction of the heat the fingers absorb is also lost through infrared emission and convection. Therefore, our additional simulations demonstrate that the ANDI-hand flux measurement includes a contribution, but not all, of the radiation absorbed by the "dummy" fingers, which explains why the simulated values are lower than the measured ones. It is worth noting that any local discrepancies have a small impact on the total absorbed radiation by the manikin, which we discuss next.

The general agreement of measured and simulated local heat fluxes allows us also to compare whole body absorbed radiation obtained with input of the same MaRTy 6-directional shortwave and longwave fluxes using the virtual manikin simulations and the standard IRM weighting method (see Section 2.5). The MRT calculated using the simulations and the IRM method differed, on average, only by 2.9°C and 3.8°C with and without considering mutual reflections for a uniformly tan-colored manikin (see **SM Fig.S14**). Since our prior work measured similar differences between MRT calculated using IRM and directly measured using a single-color (tan) physical manikin [43], the minor IRM overestimation of the MRT stems from the assumption of a cylindrical human.

### *4.2 The restricted sun conditions*

While the results of the full shade experiment demonstrated the ability of the two-color ANDI method to quantify the two spectral bands of radiation under very low shortwave illumination, the narrow urban canyon and partial shade experiments illustrated two significant limitations of the physical manikin and the hybrid MaRTy-computational approaches. In particular, in the narrow urban canyon, the differences between simulations and ANDI measurements were higher than in the unrestricted sun conditions across most zones, not just the pelvis and the hand. A closer look at the placement of ANDI and MaRTy within the narrow urban canyon in **Fig.2b** reveals the likely underlying reason for this disagreement. In particular, while the two instruments could be placed about 2 m apart without impacting each other in the unobstructed sun site (see **Fig.2a**), the combination of narrow space and complex shade in the canyon necessitated placing the biometeorological cart further from the manikin. Within this complex built environment, the radiative field changes substantially across small distances of a few meters, explaining the observed differences. For example, MaRTy is placed further from the wall, which explains why it measured a higher diffuse irradiation component that translated into elevated shortwave fluxes simulated on the back zones of the manikin. In contrast, the side and back of the physical manikin were close to the wall, reducing diffuse radiation exposure. Therefore, better results are likely to be obtained in more complex built environments if the location of the two instruments is periodically swapped to account for spatial differences across several meters. The partial shade experiment, in turn, illuminates the impacts of the irradiation heterogeneities across smaller distances.

The partial shade the Palo Verde tree provided had significant differences across distances as small as a few centimeters. It was also continually changing due to the fluttering of the branches (see **Fig.8c-d**), creating uneven exposure of the symmetric white-tan zones and of the small MaRTy sensors. If the tan zone is exposed to higher shortwave flux than the white equivalent, as in the example of the upper arm in **Fig.8d**, the $q^{"}_{i-w}$ will be lower due to the shading, leading to a higher estimate of the $S_i$ calculated using **Eq.3**, displayed in **Fig.8d**. The fluttering of the tree and high shade heterogeneities across centimeters impact the MaRTy measurements since the sensors' dimensions are about 2.5 cm. The sensor can end up being over or under-exposed depending on whether it is within the sunny or shaded "patch" and not provide representative flux measurements. Therefore, to provide accurate ANDI measurements and inputs into the simulations, the irradiation should be uniform across distance separating the furthest tan-white zones or about 0.5 m.

### *4.3 Measuring radiation attenuation provided by clothing*

The simultaneous ability to quantify shortwave and longwave irradiation and the net radiative heat flux on the manikin shell provides a unique way to assess the radiative component of various textiles' human body cooling ability in real-life conditions. The development of this method is particularly relevant as numerous new "radiatively engineered" textiles have been proposed in the literature and available on the market to improve radiative cooling [83–85,116,117]. However, the relative performance of these textiles has not been quantified in outdoor conditions with realistic human body shapes. The knowledge of the zonal shortwave and longwave fluxes, whether directionally measured using the two-color ANDI method or obtained through virtual manikin simulation using MaRTy fluxes, provides the ability to relate the spectral fabric properties to its cooling performance. As demonstrated in Section 3.4, the rudimentary model predicts $q''_{shirt}$ values within 15% of the ANDI-measured net radiative fluxes. Therefore, the introduced methodologies could be used to directly measure or theoretically estimate the cooling performance of various radiatively engineered textiles in realistic conditions.

## 5. Conclusions

In summary, we developed two novel methods for quantifying shortwave and longwave irradiation's spatial distribution on the human body. The first method employs an innovative outdoor thermal manikin featuring white-coated zones to facilitate spectral separation. In particular, by comparing measurements from symmetrically corresponding surface zones with tan and white colors, the method isolates and quantifies the contributions of shortwave and longwave radiation. The second, a hybrid experimental and computational approach, integrates radiometer measurements taken in six directions using MaRTy with virtual ANDI simulations. This technique converts the six-directional shortwave measurement into directional, reflected, and diffuse sources for the simulations. Simulation of the longwave irradiation was unnecessary, as in all tested conditions, they were nearly uniform across all the manikin zones and equal to the average of the MaRTy measurements.

We applied these methods to analyze irradiation distributions across various body parts under outdoor conditions ranging from full unobstructed sunlight to partial and complete shade in warm to extremely hot conditions. The results from the two methods were closely aligned in most cases, with notable deviations in the narrow urban canyon and partial Palo Verde tree shade that provided insights into the strengths and limitations of each approach. In particular, in very complex built environments such as the narrow canyon where radiation changes over short distances the manikin and the biometeorological cart must be sequentially switched for the results to match. The shade also needs to be homogeneous on a scale of ~0.5 m or so for the two methods to yield realistic results. It is also worth noting that in certain situations, such as with highly heated walls, cooling panels, or heating lamps, the longwave irradiation will mandate modeling as non-uniform. Similarly, the fit of the relatively simplistic direct, reflected, diffuse treatment for representing the anisotropic nature of the shortwave irradiation in situations of interest should be evaluated carefully.

Additionally, we demonstrated the utility of the manikin in quantifying the attenuation of irradiation provided by long-sleeve shirts in five different colors, ranging from white to black. In particular, wearing of the shirts reduced the heat rate absorbed by the bare upper body (i.e., tan shell) from 159W to 73.3W (white fabric), 86.7W (light blue fabric), 128W (dark blue fabric), 117W (maroon fabric), and 127W (black fabric). We also demonstrated that the ANDI-measured net radiative heat fluxes under the shirts can be predicted using a rudimentary model that considers the spectral properties of the textile and the values of the zonal shortwave and longwave irradiation fluxes. Therefore, the two approaches provide a complementary method to evaluate the performance of novel textiles with radiative functionalities in outdoor conditions with realistic human body representation. Furthermore, the detailed knowledge of shortwave and longwave irradiation across the human body in various outdoor conditions can guide the selection of fabrics with appropriate radiative functionality for the different body parts.

The introduced methodologies represent a significant advancement in quantifying the radiative part of the human-environment thermal interactions. These approaches offer the flexibility to simulate realistic irradiation scenarios for diverse built environments and varied individual characteristics. By integrating these techniques with airflow, human thermoregulation, and thermal sensation models, designers can more accurately optimize outdoor built environments for improved thermal comfort. While our outdoor thermal manikin is an expensive and one-of-a-kind instrument, its measurements agree with those obtained using the much more affordable hybrid MaRTy-simulation method. Therefore, the MaRTy instrument cart, upgraded with a 3D anemometer to account for turbulence effects in convection [43,44], can measure all inputs required to accurately simulate the 3D human heat exposure imposed by real built environments. Utilizing measured values eliminates any uncertainties stemming from assumptions in the built environment component of the coupled environment-human body-thermoregulation-thermal sensation simulations, the latest advancement in the area [86]. In contrast to MRT and thermal indices, this advanced approach can quantify thermal discomfort from localized body part exposure stemming from, for example, shortwave reflecting (e.g., cool pavements) or longwave emitting (e.g., the underside of dual-purpose photovoltaic and shading structures) engineered surfaces often used to modify built environments in extremely hot regions, thereby broadening the simulation applicability and relevance to research and practice.

**Acknowledgments**

This research was supported by the National Science Foundation Leading Engineering for America's Prosperity, Health, and Infrastructure (LEAP HI) #2152468 award, while the thermal manikin was funded by a National Science Foundation Major Research Instrumentation grant #2117917. Dr. Sylwester Wereski's participation in the research was possible thanks to funding from the National Science Center in Poland under the MINIATURA 5 grant "The occurrence of strong heat stress in hot and dry climate condition" (2021/05/X/ST10/00587). The authors gratefully acknowledge the use of spectroscopy equipment within the Eyring Center for Solid State Science at Arizona State University.

**Declaration of competing interest**

The authors declare that they have no known competing financial interests or personal relationships that could have appeared to influence the work reported in this paper.

**Author contributions**

KR: conceptualization, experiments, analysis, funding; AJ, SHV, LB, SW, KS, CTJ, MAb: field experiments and analysis; GM: textile suit preparation; AM: conceptualization, funding; All authors contributed to manuscript writing and editing.

**Declaration of generative AI and AI-assisted technologies in the writing process**

During the preparation of this work the author(s) used ChatGPT to improve the manuscript draft. After using this tool/service, the author(s) reviewed and edited the content as needed and take(s) full responsibility for the content of the publication.

**References**

[1] S.E. Perkins, L. V Alexander, J.R. Nairn, Increasing frequency, intensity and duration of observed global heatwaves and warm spells, Geophys Res Lett 39 (2012).

[2] C.M. Powis, D. Byrne, Z. Zobel, K.N. Gassert, A.C. Lute, C.R. Schwalm, Observational and model evidence together support wide-spread exposure to noncompensable heat under continued global warming, Sci Adv 9 (2023) eadg9297.

[3] K.L. Ebi, A. Capon, P. Berry, C. Broderick, R. de Dear, G. Havenith, Y. Honda, R.S. Kovats, W. Ma, A. Malik, Hot weather and heat extremes: health risks, The Lancet 398 (2021) 698–708.

[4] K.L. Ebi, J. Vanos, J.W. Baldwin, J.E. Bell, D.M. Hondula, N.A. Errett, K. Hayes, C.E. Reid, S. Saha, J. Spector, Extreme weather and climate change: population health and health system implications, Annu Rev Public Health 42 (2021) 293.

[5] J. Vanos, G. Guzman-Echavarria, J.W. Baldwin, C. Bongers, K.L. Ebi, O. Jay, A physiological approach for assessing human survivability and liveability to heat in a changing climate, Nat Commun 14 (2023) 7653. https://doi.org/10.1038/s41467-023-43121-5.

[6] A.D. Flouris, P.C. Dinas, L.G. Ioannou, L. Nybo, G. Havenith, G.P. Kenny, T. Kjellstrom, Workers' health and productivity under occupational heat strain: a systematic review and meta-analysis, Lancet Planet Health 2 (2018) e521–e531.

[7] T. Kjellstrom, N. Maître, C. Saget, O. Matthias, T. Karimova, International Labour Organization: Working on a warmer planet: The impact of heat stress on labour

productivity and decent work, (2019). https://www.ilo.org/global/publications/books/WCMS_711919/lang--en/index.htm.
[8] T. Kjellstrom, R.S. Kovats, S.J. Lloyd, T. Holt, R.S.J. Tol, The direct impact of climate change on regional labor productivity, Arch Environ Occup Health 64 (2009) 217–227.
[9] O. Jay, A. Capon, P. Berry, C. Broderick, R. de Dear, G. Havenith, Y. Honda, R.S. Kovats, W. Ma, A. Malik, Reducing the health effects of hot weather and heat extremes: from personal cooling strategies to green cities, The Lancet 398 (2021) 709–724.
[10] N. Nazarian, E.S. Krayenhoff, B. Bechtel, D.M. Hondula, R. Paolini, J. Vanos, T. Cheung, W.T.L. Chow, R. de Dear, O. Jay, Integrated assessment of urban overheating impacts on human life, Earths Future 10 (2022) e2022EF002682.
[11] D. Lai, W. Liu, T. Gan, K. Liu, Q. Chen, A review of mitigating strategies to improve the thermal environment and thermal comfort in urban outdoor spaces, Science of the Total Environment 661 (2019) 337–353.
[12] S. Yang, L.L. Wang, T. Stathopoulos, A.M. Marey, Urban microclimate and its impact on built environment–a review, Build Environ 238 (2023) 110334.
[13] E.S. Krayenhoff, A.M. Broadbent, L. Zhao, M. Georgescu, A. Middel, J.A. Voogt, A. Martilli, D.J. Sailor, E. Erell, Cooling hot cities: a systematic and critical review of the numerical modelling literature, Environmental Research Letters 16 (2021) 053007.
[14] Z. Ren, Y. Fu, Y. Dong, P. Zhang, X. He, Rapid urbanization and climate change significantly contribute to worsening urban human thermal comfort: A national 183-city, 26-year study in China, Urban Clim 43 (2022) 101154.
[15] D. Lai, Z. Lian, W. Liu, C. Guo, W. Liu, K. Liu, Q. Chen, A comprehensive review of thermal comfort studies in urban open spaces, Science of the Total Environment 742 (2020) 140092.
[16] R. Abd Elraouf, A. Elmokadem, N. Megahed, O.A. Eleinen, S. Eltarabily, The impact of urban geometry on outdoor thermal comfort in a hot-humid climate, Build Environ 225 (2022) 109632.
[17] M. Nikolopoulou, S. Lykoudis, Thermal comfort in outdoor urban spaces: Analysis across different European countries, Build Environ 41 (2006) 1455–1470. https://doi.org/https://doi.org/10.1016/j.buildenv.2005.05.031.
[18] M. Taleghani, Outdoor thermal comfort by different heat mitigation strategies-A review, Renewable and Sustainable Energy Reviews 81 (2018) 2011–2018.
[19] V.K. Turner, A. Middel, J.K. Vanos, Shade is an essential solution for hotter cities, Nature 619 (2023) 694–697.
[20] Y. Yu, R. de Dear, Thermal respite for pedestrians in overheated urban environments – Introduction of a dynamic analysis of outdoor thermal comfort, Sustain Cities Soc 86 (2022) 104149. https://doi.org/https://doi.org/10.1016/j.scs.2022.104149.
[21] R.F. Rupp, N.G. Vásquez, R. Lamberts, A review of human thermal comfort in the built environment, Energy Build 105 (2015) 178–205.
[22] Q. Zhao, Z. Lian, D. Lai, Thermal comfort models and their developments: A review, Energy and Built Environment 2 (2021) 21–33.

[23] D. Lai, Z. Lian, W. Liu, C. Guo, W. Liu, K. Liu, Q. Chen, A comprehensive review of thermal comfort studies in urban open spaces, Science of the Total Environment 742 (2020) 140092.
[24] F. Binarti, M.D. Koerniawan, S. Triyadi, S.S. Utami, A. Matzarakis, A review of outdoor thermal comfort indices and neutral ranges for hot-humid regions, Urban Clim 31 (2020) 100531.
[25] Y. Dzyuban, G.N.Y. Ching, S.K. Yik, A.J. Tan, S. Banerjee, P.J. Crank, W.T.L. Chow, Outdoor thermal comfort research in transient conditions: A narrative literature review, Landsc Urban Plan 226 (2022) 104496.
[26] S. Shooshtarian, C.K.C. Lam, I. Kenawy, Outdoor thermal comfort assessment: A review on thermal comfort research in Australia, Build Environ 177 (2020) 106917.
[27] P. Kumar, A. Sharma, Study on importance, procedure, and scope of outdoor thermal comfort–A review, Sustain Cities Soc 61 (2020) 102297.
[28] J. Spagnolo, R.J. de Dear, A field study of thermal comfort in outdoor and semi-outdoor environments in subtropical {S}ydney {A}ustralia, Build Environ 38 (2003) 721–738.
[29] M. Nakayoshi, M. Kanda, R. Shi, R. de Dear, Outdoor thermal physiology along human pathways: a study using a wearable measurement system, Int J Biometeorol 59 (2015) 503–515.
[30] C. Vasilikou, M. Nikolopoulou, Outdoor thermal comfort for pedestrians in movement: thermal walks in complex urban morphology, Int J Biometeorol 64 (2020) 277–291.
[31] A. Middel, N. Selover, B. Hagen, N. Chhetri, Impact of shade on outdoor thermal comfort—a seasonal field study in Tempe, Arizona, Int J Biometeorol 60 (2016) 1849–1861.
[32] C.R. de Freitas, E.A. Grigorieva, A comprehensive catalogue and classification of human thermal climate indices, Int J Biometeorol 59 (2015) 109–120.
[33] S. Thorsson, T. Honjo, F. Lindberg, I. Eliasson, E.-M. Lim, Thermal comfort and outdoor activity in Japanese urban public places, Environ Behav 39 (2007) 660–684.
[34] M. Nikolopoulou, N. Baker, K. Steemers, Thermal comfort in outdoor urban spaces: understanding the human parameter, Solar Energy 70 (2001) 227–235.
[35] R. Aghamolaei, M.M. Azizi, B. Aminzadeh, J. O'Donnell, A comprehensive review of outdoor thermal comfort in urban areas: Effective parameters and approaches, Energy & Environment 34 (2023) 2204–2227.
[36] J. Pickup, R. de Dear, An outdoor thermal comfort index (OUT_SET*)-part I-the model and its assumptions, in: Biometeorology and Urban Climatology at the Turn of the Millenium. Selected Papers from the Conference ICB-ICUC, 2000: pp. 279–283.
[37] T. Huang, J. Li, Y. Xie, J. Niu, C.M. Mak, Simultaneous environmental parameter monitoring and human subject survey regarding outdoor thermal comfort and its modelling, Build Environ 125 (2017) 502–514. https://doi.org/https://doi.org/10.1016/j.buildenv.2017.09.015.
[38] J. Zou, Y. Yu, J. Liu, J. Niu, K. Chauhan, C. Lei, Field measurement of the urban pedestrian level wind turbulence, Build Environ 194 (2021) 107713.

[39] Y. Yu, R. de Dear, K. Chauhan, J. Niu, Impact of wind turbulence on thermal perception in the urban microclimate, Journal of Wind Engineering and Industrial Aerodynamics 216 (2021) 104714.
[40] J. Zou, J. Liu, J. Niu, Y. Yu, C. Lei, Convective heat loss from computational thermal manikin subject to outdoor wind environments, Build Environ (2020) 107469.
[41] Y. Yu, J. Liu, K. Chauhan, R. de Dear, J. Niu, Experimental study on convective heat transfer coefficients for the human body exposed to turbulent wind conditions, Build Environ 169 (2020) 106533.
[42] S. Zhou, J. Niu, Measurement of the convective heat transfer coefficient of the human body in the lift-up design, in: E3S Web of Conferences, EDP Sciences, 2022: p. 03001.
[43] A. Joshi, S.H. Viswanathan, A.K. Jaiswal, K. Sadeghi, L. Bartels, R.M. Jain, G. Pathikonda, J.K. Vanos, A. Middel, K. Rykaczewski, Characterization of human extreme heat exposure using an outdoor thermal manikin, Science of the Total Environment 923 (2024) 171525.
[44] S. Zhou, Y. Yu, K.C.S. Kwok, J. Niu, Onsite measurements of pedestrian-level wind and preliminary assessment of effects of turbulence characteristics on human body convective heat transfer, Energy Build (2024) 114448.
[45] I.M.S. Abouelhamd, K. Kuga, K. Ito, Convective heat transfer and drag coefficients of human body in multiple crowd densities and configurations in semi-outdoor scenarios, Build Environ 265 (2024) 111983.
[46] K. Blazejczyk, H. Nilsson, I. Holmér, Solar heat load on man, Int J Biometeorol 37 (1993) 125–132.
[47] C.R. Underwood, E.J. Ward, The solar radiation area of man, Ergonomics 9 (1966) 155–168.
[48] K. Kubaha, D. Fiala, J. Toftum, A.H. Taki, Human projected area factors for detailed direct and diffuse solar radiation analysis, Int J Biometeorol 49 (2004) 113–129.
[49] A. Zani, H.D. Richardson, A. Tono, S. Schiavon, E. Arens, A simulation-based design analysis for the assessment of indoor comfort under the effect of solar radiation, (2019).
[50] Y. Kurazumi, T. Horikoshi, K. Hirayama, T. Tsuchikawa, Y. Kobayashi, The influence of asymmetric and uneven thermal radiation environments upon the human body, In the case of constant operative temperature, Journal of Architecture, Planning and Environmental Engineering 447 (1993) 17–26.
[51] Y. Kurazumi, K. Saito, T. Horikoshi, The influence of asymmetric thermal radiation environments upon the human body, In the case of constant operative temperature and right and left, back and forth asymmetry, Japanese Journal of Biometeorology 31 (1994) 75–84.
[52] Y. Kurazumi, K. Fukagawa, T. Sakoi, A. Aruninta, E. Kondo, K. Yamashita, Skin temperature and body surface section in non-uniform and asymmetric outdoor thermal environment, Health N Hav 10 (2018) 1321.
[53] K. Kubaha, Asymmetric radiant fields and human thermal comfort, (2005).
[54] K. Parsons, Human thermal environments: the effects of hot, moderate, and cold environments on human health, comfort, and performance, CRC press, 2014.

[55] K. Parsons, Human heat stress, CRC Press, 2019.
[56] N. Kántor, J. Unger, The most problematic variable in the course of human-biometeorological comfort assessment—the mean radiant temperature, Central European Journal of Geosciences 3 (2011) 90–100.
[57] D. Aviv, H. Guo, A. Middel, F. Meggers, Evaluating radiant heat in an outdoor urban environment: Resolving spatial and temporal variations with two sensing platforms and data-driven simulation, Urban Clim 35 (2021) 100745.
[58] E.L. Krüger, F.O. Minella, A. Matzarakis, Comparison of different methods of estimating the mean radiant temperature in outdoor thermal comfort studies, Int J Biometeorol 58 (2014) 1727–1737.
[59] S. Thorsson, F. Lindberg, I. Eliasson, B. Holmer, Different methods for estimating the mean radiant temperature in an outdoor urban setting, Int J Climatol 27 (2007) 1983–1993.
[60] H. Guo, D. Aviv, M. Loyola, E. Teitelbaum, N. Houchois, F. Meggers, On the understanding of the mean radiant temperature within both the indoor and outdoor environment, a critical review, Renewable and Sustainable Energy Reviews 117 (2020) 109207.
[61] M. Nakayoshi, M. Kanda, R. de Dear, Globe Anemo-radiometer, Boundary Layer Meteorol 155 (2015) 209–227.
[62] J.K. Vanos, K. Rykaczewski, A. Middel, D.J. Vecellio, R.D. Brown, T.J. Gillespie, Improved methods for estimating mean radiant temperature in hot and sunny outdoor settings, Int J Biometeorol 65 (2021) 967–983.
[63] K. Rykaczewski, A. Joshi, S.H. Viswanathan, S.S. Guddanti, K. Sadeghi, M. Gupta, A.K. Jaiswal, K. Kompally, G. Pathikonda, R. Barlett, J.K. Vanos, A. Middel, A simple three-cylinder radiometer and low-speed anemometer to characterize human extreme heat exposure, Int J Biometeorol 68 (2024) 1081–1092.
[64] A. Middel, E.S. Krayenhoff, Micrometeorological determinants of pedestrian thermal exposure during record-breaking heat in Tempe, Arizona: Introducing the MaRTy observational platform, Science of the Total Environment 687 (2019) 137–151.
[65] P. Höppe, Ein neues Verfahren zur Bestimmung der mittleren Strahlungstemperatur im Freien, Wetter Und Leben 44 (1992) 147–151.
[66] A. Middel, M. Huff, E.S. Krayenhoff, A. Udupa, F.A. Schneider, PanoMRT: Panoramic infrared thermography to model human thermal exposure and comfort, Science of The Total Environment 859 (2023) 160301. https://doi.org/https://doi.org/10.1016/j.scitotenv.2022.160301.
[67] J.A.P. y Miño, C. Lawrence, B. Beckers, Visual metering of the urban radiative environment through 4π imagery, Infrared Phys Technol 110 (2020) 103463.
[68] J.A.P. y Miño, N. Duport, B. Beckers, Pixel-by-pixel rectification of urban perspective thermography, Remote Sens Environ 266 (2021) 112689.
[69] T. Sakoi, K. Tsuzuki, S. Kato, R. Ooka, D. Song, S. Zhu, Thermal comfort, skin temperature distribution, and sensible heat loss distribution in the sitting posture in various asymmetric radiant fields, Build Environ 42 (2007) 3984–3999. https://doi.org/https://doi.org/10.1016/j.buildenv.2006.10.050.

[70] E. Halawa, J. van Hoof, V. Soebarto, The impacts of the thermal radiation field on thermal comfort, energy consumption and control—A critical overview, Renewable and Sustainable Energy Reviews 37 (2014) 907–918. https://doi.org/https://doi.org/10.1016/j.rser.2014.05.040.
[71] P.O. Fanger, Thermal Comfort, McGraw-Hill Book Company, New York, 1972.
[72] F. Kalmár, Impact of elevated air velocity on subjective thermal comfort sensation under asymmetric radiation and variable airflow direction, J Build Phys 42 (2017) 173–193. https://doi.org/10.1177/1744259117737783.
[73] H. Zhao, S. Wang, Y. Zhang, L. Zhao, Y. Zhai, R.D. Brown, L. Jin, R. Wu, The effect of solar radiation on pedestrian thermal comfort: A climate chamber experiment, Build Environ 245 (2023) 110869.
[74] B.R. Anupam, U.C. Sahoo, A.K. Chandrappa, P. Rath, Emerging technologies in cool pavements: A review, Constr Build Mater 299 (2021) 123892.
[75] M. Santamouris, Using cool pavements as a mitigation strategy to fight urban heat island—A review of the actual developments, Renewable and Sustainable Energy Reviews 26 (2013) 224–240.
[76] Y. Qin, A review on the development of cool pavements to mitigate urban heat island effect, Renewable and Sustainable Energy Reviews 52 (2015) 445–459.
[77] J. Anand, D.J. Sailor, Role of pavement radiative and thermal properties in reducing excess heat in cities, Solar Energy 242 (2022) 413–423.
[78] F.A. Schneider, J.C. Ortiz, J.K. Vanos, D.J. Sailor, A. Middel, Evidence-based guidance on reflective pavement for urban heat mitigation in Arizona, Nat Commun 14 (2023) 1467.
[79] A. Middel, V.K. Turner, F.A. Schneider, Y. Zhang, M. Stiller, Solar reflective pavements—A policy panacea to heat mitigation?, Environmental Research Letters 15 (2020) 64016.
[80] M. Migliari, J. Despax, L. Chesne, O. Baverel, Street albedos repartition's effects on urban heat island and outdoor thermal comfort, in: 6th International Conference on Countermeasures to Urban Heat Islands (IC2UHI), 2023: pp. 1–10.
[81] E. Erell, D. Pearlmutter, D. Boneh, P.B. Kutiel, Effect of high-albedo materials on pedestrian heat stress in urban street canyons, Urban Clim 10 (2014) 367–386.
[82] Y. Nakamura, Y. Asano, A. Suzuki-Parker, H. Kusaka, Verification of heat stress mitigation effects by UV parasols using UTCI observations and thermal sensory questionnaire survey, Build Environ 266 (2024) 112025.
[83] L. Cai, Y. Peng, J. Xu, C. Zhou, C. Zhou, P. Wu, D. Lin, S. Fan, Y. Cui, Temperature regulation in colored infrared-transparent polyethylene textiles, Joule 3 (2019) 1478–1486.
[84] L. Cai, A.Y. Song, W. Li, P. Hsu, D. Lin, P.B. Catrysse, Y. Liu, Y. Peng, J. Chen, H. Wang, Spectrally selective nanocomposite textile for outdoor personal cooling, Advanced Materials 30 (2018) 1802152.
[85] J.K. Tong, X. Huang, S. v Boriskina, J. Loomis, Y. Xu, G. Chen, Infrared-transparent visible-opaque fabrics for wearable personal thermal management, ACS Photonics 2 (2015) 769–778.

[86] K. Dharmasastha, H. Liang, J. Lin, Y. Xie, Y. Yu, J. Niu, Evaluating thermal sensation in outdoor environments - Different methods of coupling CFD and radiation modelling with a human body thermoregulation model, Build Environ 266 (2024) 112081. https://doi.org/https://doi.org/10.1016/j.buildenv.2024.112081.

[87] F. Meggers, H. Guo, E. Teitelbaum, G. Aschwanden, J. Read, N. Houchois, J. Pantelic, E. Calabrò, The Thermoheliodome–"Air conditioning" without conditioning the air, using radiant cooling and indirect evaporation, Energy Build 157 (2017) 11–19.

[88] K. Rykaczewski, J.K. Vanos, A. Middel, Anisotropic radiation source models for computational thermal manikin simulations based on common radiation field measurements, Build Environ 208 (2022) 108636.

[89] K. Kubaha, D. Fiala, K.J. Lomas, Predicting human geometry-related factors for detailed radiation analysis in indoor spaces, in: Eighth International IBPSA Conference, © IBPSA, Eindhoven, Netherlands, 2003: pp. 681–688.

[90] S. Park, S.E. Tuller, Comparison of human radiation exchange models in outdoor areas, Theor Appl Climatol 105 (2011) 357–370.

[91] K. Rykaczewski, L. Bartels, D.M. Martinez, S.H. Viswanathan, Human body radiation area factors for diverse adult population, Int J Biometeorol 66 (2022) 2357–2367.

[92] N. Youssef, K. D'Avignon, Investigation into the Pertinence of Using Child-Specific Radiation Data for Thermal Comfort Calculations, ASHRAE Trans 129 (2023) 81+. https://link.gale.com/apps/doc/A787006209/AONE?u=anon~51d6ac74&sid=googleScholar&xid=1aaa4084.

[93] M.F.B. Suhaimi, W.G. Kim, C.-W. Cho, H. Lee, Impact of solar radiation on human comfort in a vehicle cabin: An analysis of body segment mean radiant temperature, Build Environ 245 (2023) 110849.

[94] A. Matzarakis, F. Rutz, H. Mayer, Modelling radiation fluxes in simple and complex environments: basics of the RayMan model, Int J Biometeorol 54 (2010) 131–139.

[95] F. Lindberg, B. Holmer, S. Thorsson, SOLWEIG 1.0–Modelling spatial variations of 3D radiant fluxes and mean radiant temperature in complex urban settings, Int J Biometeorol 52 (2008) 697–713.

[96] C. V Gál, N. Kántor, Modeling mean radiant temperature in outdoor spaces, A comparative numerical simulation and validation study, Urban Clim 32 (2020) 100571.

[97] S.J. Rees, K.J. Lomas, D. Fiala, Predicting local thermal discomfort adjacent to glazing, ASHRAE Trans 114 (2008) 1–10.

[98] D. Aviv, M. Hou, E. Teitelbaum, H. Guo, F. Meggers, Simulating Invisible Light: Adapting Lighting and Geometry Models for Radiant Heat Transfer, in: Proc. Symp. Simul. Archit. Urban Des, 2020: pp. 311–318.

[99] S. Murakami, S. Kato, J. Zeng, Combined simulation of airflow, radiation and moisture transport for heat release from a human body, Build Environ 35 (2000) 489–500.

[100] R. Streblow, D. Müller, I. Gores, P. Bendfeldt, A coupled simulation of the thermal environment and thermal comfort with an adapted Tanabe comfort model, in: 7th International Thermal Manikin and Modelling Meeting, Portugal, Http://Www. Adai. Pt/7i3/Paper. Html, 2008.

[101] S. Tanabe, K. Kobayashi, J. Nakano, Y. Ozeki, M. Konishi, Evaluation of thermal comfort using combined multi-node thermoregulation (65MN) and radiation models and computational fluid dynamics (CFD), Energy Build 34 (2002) 637–646.
[102] S. Zhu, S. Kato, R. Ooka, T. Sakoi, K. Tsuzuki, Development of a computational thermal manikin applicable in a non-uniform thermal environment—Part 2: Coupled simulation using Sakoi's human thermal physiological model, HVAC&R Res 14 (2008) 545–564.
[103] R. Yousaf, D. Fiala, A. Wagner, Numerical simulation of human radiation heat transfer using a mathematical model of human physiology and computational fluid dynamics (CFD), in: High Performance Computing in Science and Engineering07, Springer, 2008: pp. 647–666.
[104] P.C. Cropper, T. Yang, M. Cook, D. Fiala, R. Yousaf, Coupling a model of human thermoregulation with computational fluid dynamics for predicting human–environment interaction, J Build Perform Simul 3 (2010) 233–243. https://doi.org/10.1080/19401491003615669.
[105] L. Schellen, M. Loomans, B.R.M. Kingma, M.H. De Wit, A.J.H. Frijns, W.D. van Marken Lichtenbelt, The use of a thermophysiological model in the built environment to predict thermal sensation: coupling with the indoor environment and thermal sensation, Build Environ 59 (2013) 10–22.
[106] B. Choudhary, A coupled CFD-thermoregulation model for air ventilation clothing, Energy Build 268 (2022) 112206.
[107] B. Choudhary, F. Wang, Y. Ke, J. Yang, Development and experimental validation of a 3D numerical model based on CFD of the human torso wearing air ventilation clothing, Int J Heat Mass Transf 147 (2020) 118973.
[108] J. Yang, S. Ni, W. Weng, Modelling heat transfer and physiological responses of unclothed human body in hot environment by coupling CFD simulation with thermal model, International Journal of Thermal Sciences 120 (2017) 437–445.
[109] B.A. MacRae, C.M. Spengler, A. Psikuta, R.M. Rossi, S. Annaheim, A thermal skin model for comparing contact skin temperature sensors and assessing measurement errors, Sensors 21 (2021) 4906.
[110] K. Rykaczewski, T. Dhanote, Analysis of thermocouple-based finger contact temperature measurements, J Therm Biol (2022) 103293.
[111] A. Joshi, L. Bartels, S.H. Viswanathan, D.M. Martinez, K. Sadeghi, A.K. Jaiswal, D. Collins, K. Rykaczewski, Evaluation of thermal properties and thermoregulatory impacts of lower back exosuit using thermal manikin, Int J Ind Ergon 98 (2023) 103517.
[112] B. Holmer, F. Lindberg, D. Rayner, S. Thorsson, How to transform the standing man from a box to a cylinder–a modified methodology to calculate mean radiant temperature in field studies and models, in: Proceedings of the 9th International Conference on Urban Climate (ICUC9), Toulouse, France, 2015: pp. 20–24.
[113] R. Perez, R. Seals, P. Ineichen, R. Stewart, D. Menicucci, A new simplified version of the Perez diffuse irradiance model for tilted surfaces, Solar Energy 39 (1987) 221–231.

[114] E.H. Wissler, Human Temperature Control: A quantitative Approach, 1st ed., Springer, Berlin, Germany, 2018.
[115] S.H. Viswanathan, A. Joshi, L. Bartels, K. Sadeghi, J.K. Vanos, K. Rykaczewski, Impact of Human Body Shape on Free Convection Heat Transfer, PLoS One (2025).
[116] E.D. Patamia, M.K. Yee, T.L. Andrew, Microstructured Reflective Coatings on Commodity Textiles for Passive Personal Cooling, ACS Appl Mater Interfaces 16 (2024) 59424–59433.
[117] F. Zhu, Q.Q. Feng, Recent advances in textile materials for personal radiative thermal management in indoor and outdoor environments, International Journal of Thermal Sciences 165 (2021) 106899.

# Supplemental Material for “Resolving shortwave and longwave irradiation distributions across the human body in outdoor built environments”

*Kambiz Sadeghi,[1,2] Shri H. Viswanathan,[1] Ankit Joshi,[1,2] Lyle Bartels,[1] Sylwester Wereski,[3] Cibin T. Jose, [1] Galina Mihaleva,[4] Muhammad Abdullah,[5] Ariane Middel,[2,6,7] and Konrad Rykaczewski,[1,2*]*

1. *School for Engineering of Matter, Transport and Energy, Arizona State University, Tempe, AZ 85282, USA*
2. *Julie Ann Wrigley Global Futures Laboratory, Arizona State University, Tempe, AZ 85282, USA*
3. *Institute of Earth and Environmental Sciences, Maria Curie-Skłodowska University, Lublin, Poland*
4. *Arizona State University Fashion Institute of Design and Merchandizing (ASU FIDM), Herberger Institute for Design and the Arts, Arizona State University, Tempe, AZ 85282, USA*
5. *School of Sustainability, Arizona State University, Tempe, AZ 85282, USA*
6. *School of Arts, Media and Engineering, Herberger Institute for Design and the Arts, Arizona State University, Tempe, AZ 85282, USA*
7. *School of Computing and Augmented Intelligence, Arizona State University, Tempe, AZ 85282, USA*

**corresponding author email: konradr@asu.edu, phone: (480) 965-4912, address: Arizona State University, 501 E Tyler Mall, Tempe, AZ 85282, USA*

**S1. Spectral absorptivity of the white and tan coatings**

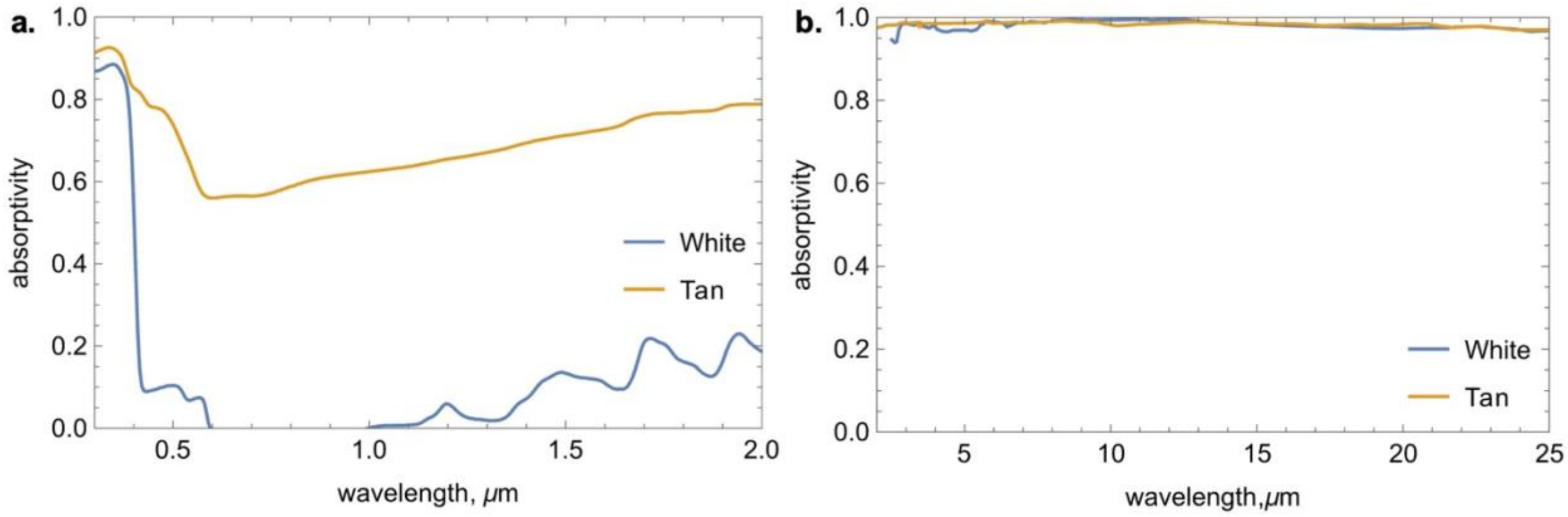

**Fig.S1 a.** Shortwave and **b.** longwave hemispherical spectral absorptivity of the white and tan coatings measured using UV-VIS and FTIR (see procedure in Rykaczewski et al.[1]).

**S2. Thermal resistance of the white paint**

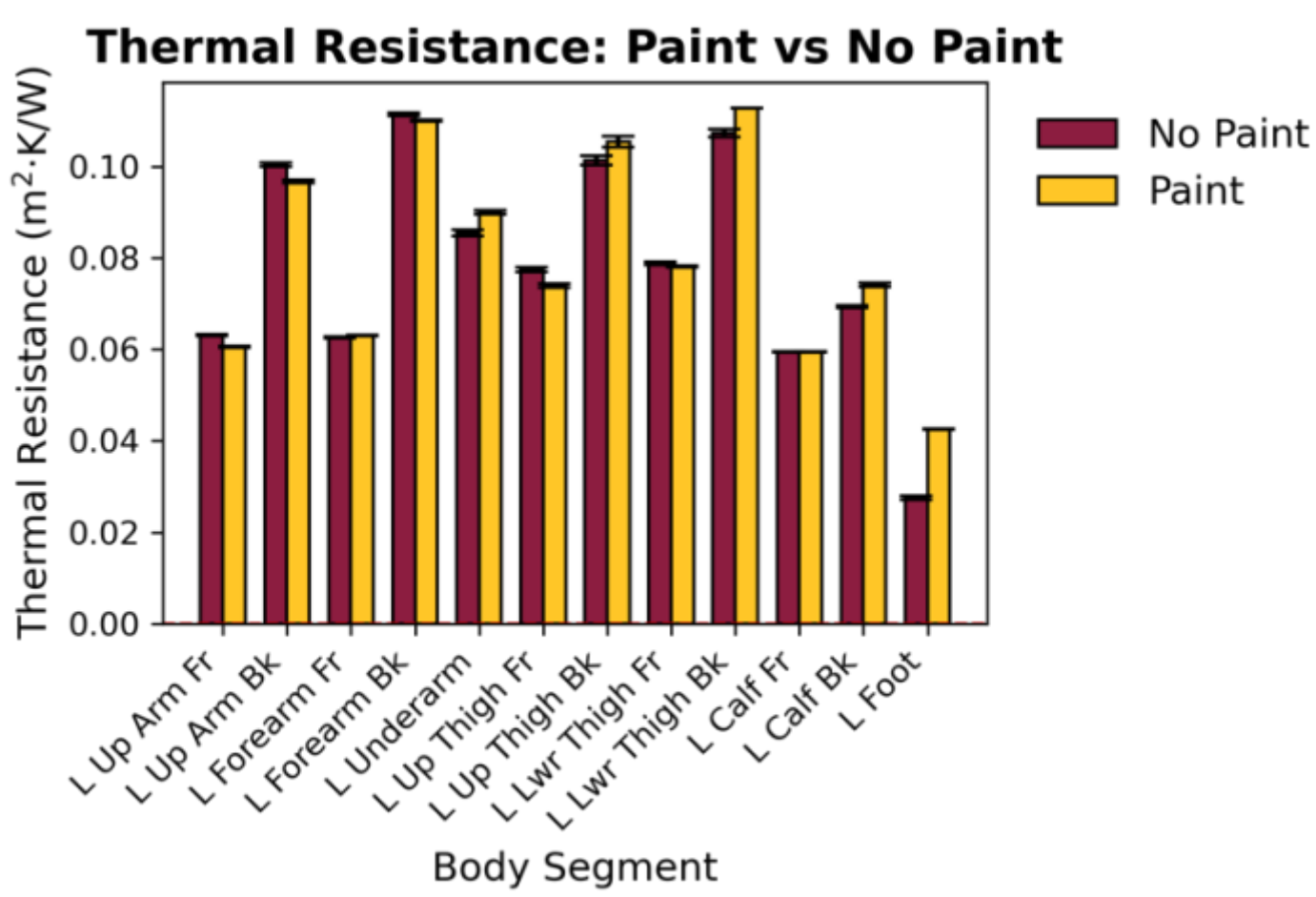

**Fig.S2** Comparison of the thermal resistance measured by the manikin on zones with and without the thin layer of white paint conducted in a climatic chamber and wind enclosure in accordance with the ASTM Standard F1291 [2]. During the measurements, the ambient conditions were carefully controlled, maintaining a temperature of 23 ± 0.7°C, a relative humidity of 50 ± 3%, and an airspeed of 0.4 ± 0.05 $m \cdot s^{-1}$.

**S3. The white onesie based method to resolve shortwave and longwave irradiation**

The sequential variation of the two color ANDI method relied on measurements conducted with the manikin nude and dressed in the custom-made white onesie. The fabric has a shortwave and longwave absorptivities of 0.20 and 0.97, with spectral distribution shown in **Fig.S3a**. Naturally, the fabric adds a thermal resistance to the manikin, requiring subcooling the shell below the air temperature to achieve exterior fabric temperature equal to that of air. The ASTM F1291 standard test method for measuring thermal insulation of clothing repeated 3 times to measure $R_{fab}$ (see **Fig.S3b** for images), with values shown in **Fig.S3c**. Without any corrections, the fabric resistance contributed to decrease of the shell temperature of 35°C to about, on average, 31.5°C on the exterior of the fabric in the ASTM test conditions (see example infrared images in **Fig.S3d**). Accordingly, we made a modification to the manikin control software to account for these resistances. In particular, using one dimensional thermal circuit, the steady state temperature on the shell must can be related to that of required fabric exterior as:

$$T_{sk} = \frac{T_{in} + \frac{R_{sens}}{R_{fab}} T_{fab}}{1 + \frac{R_{sens}}{R_{fab}}} \quad \text{(S.1)}$$

Where $T_{sk}$ is the manikin skin temperature (°C), $R_{sens}$ is the thermal resistance of the manikin shell (°C*m²/W) which is provided by the manufacturer, $R_{fab}$ is the thermal resistance added by the fabric (°C*m²/W), and $T_{fab}$ is the desired temperature of the outer surface of the fabric (°C). In this case $T_{fab}$ should be equal to the ambient air temperature. **Fig.S3e** shows typical cooling curve for the fabric under intense mimicked solar radiation shown in main manuscript. Evidently, the 20 to 30 minute settling period, in addition to 10 to 15 minute dressing/undressing time, made the process too slow for outdoor measurements.

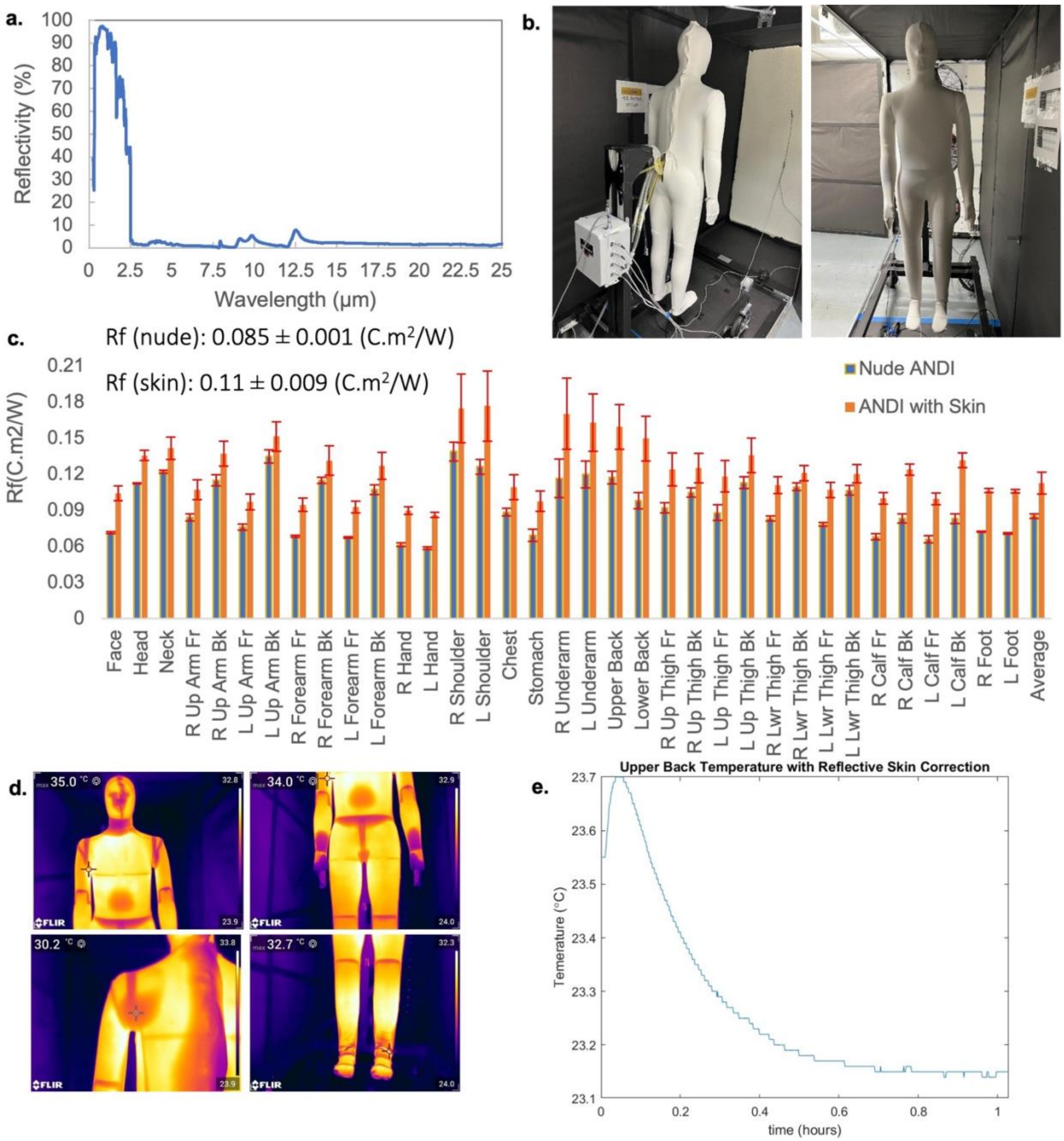

**Fig.S3 a.** the spectral reflectivity of the white fabric used to make the onesie, **b.** images of the thermal manikin wearing the onesie in the wind tunnel used to measure c. thermal resistances of each zone with and without the onesie, d. infrared images of the manikin without the temperature compensation showning non-uniformities, and e. settling time under intense mimicked solar radiation (see Fig.3b).

## S4. Spectral, hemispherical absorptivity of the fabrics

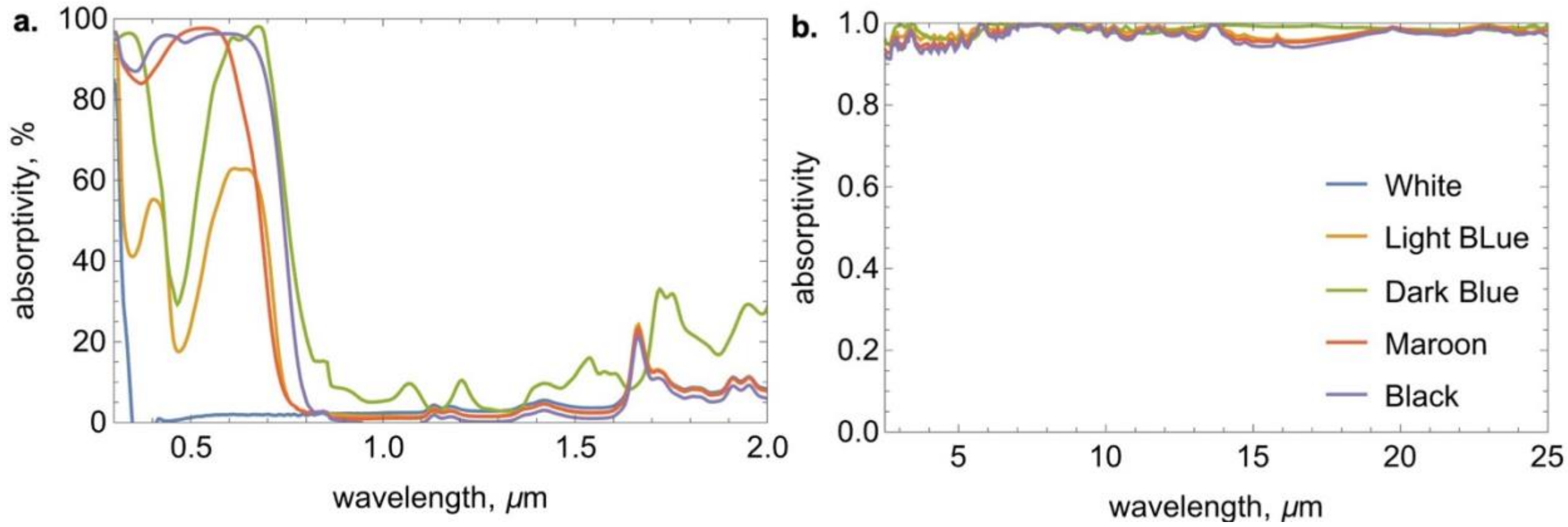

**Fig.S4 a.** shortwave hemispherical absorptivity (100-transmissivity-absorptivity) and **b.** longwave hemispherical spectral absorptivity of the white, light blue, dark blue, maroon, and black textiles measured using UV-VIS and FTIR (see procedure in Rykaczewski et al.[1]).

## S5. Directional reflectivity measurements

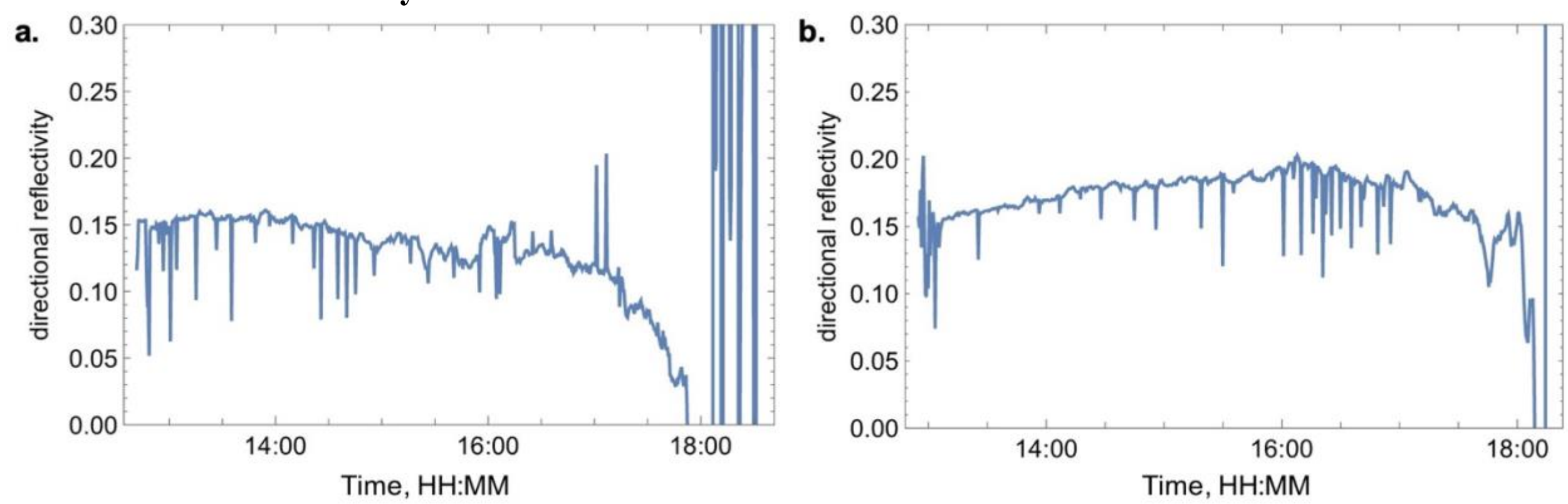

**Fig.S5** The directional reflectivity calculated using Eq.7 using MaRTy measurements conducted on **a.** September 11th and **b.** September 13th of 2024.

## S6. Quantifying MaRTy misalignment from cardinal directions

Following Holmer [3], we calculate the potential misalignment angle between MaRTy and cardinal directions (Δ, see **Fig.S6a**) from the cross-over of the east and west facing sensors in terms of the azimuth angle. We illustrate the process with full day fluxes collected on August 15th, 2024 in the same location as the ANDI radiation experiments (see **Fig.S6b**). We note that in August the azimuth angle at sunrise is around 70° (see **Fig.S6c**), therefore the north-facing sensor is exposed to direct solar radiation. However, the asymmetry between the north- and south-facing signals as well as that between the east and west facing sensors implies that MaRTy was misaligned. When plotted as function of the azimuth angle, the difference between azimuth of 180° (i.e., solar noon) and the angle for cross-over the east- and west-facing is clearly evident (see **Fig.S6d**), particularly when the diffuse solar radiation calculated using Eq.5 is subtracted (see **Fig.S6e**). As illustrated in **Fig.S6e,** Δ can be readily obtained from this plot. In case of the experiment was started after solar noon, **Fig.S6f** shows that Δ can also be estimated using this method because $S_{west} - S_{diff}$ is nearly linear near solar noon and can be extrapolated.

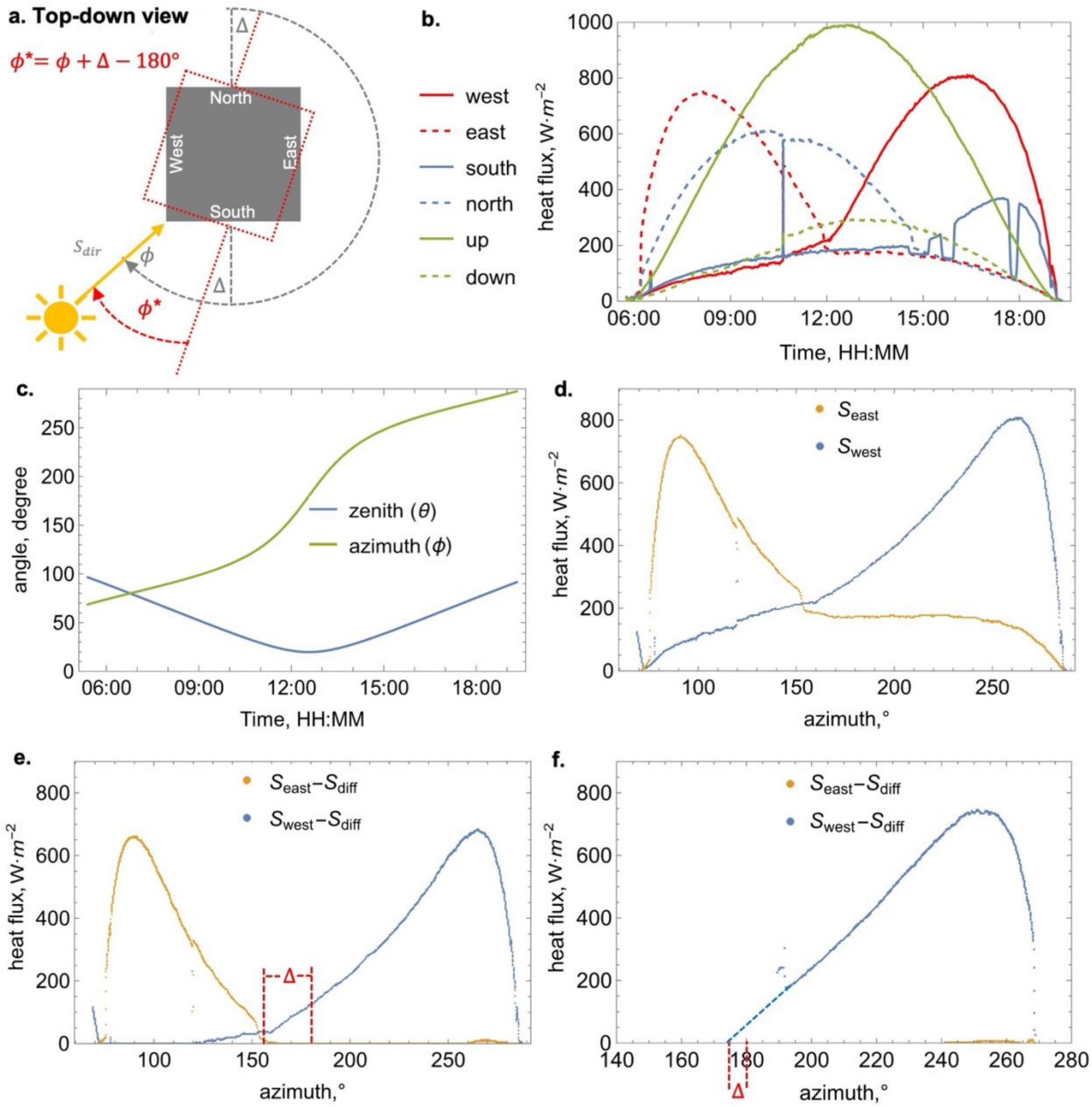

**Fig.S6 a.** Illustration of the MaRTy misalignment angle Δ and the local azimuth angle $\phi^*$, **b.** the 6-directional shortwave fluxes measured on the top floor of the Novus parking lot on August 15th, 2024 and **c.** the corresponding solar angles, and **d.** $S_{east}$ and $S_{west}$ and **e.** $S_{east} - S_{diff}$ and $S_{west} - S_{diff}$ fluxes from which Δ can be determined; **f.** $S_{east} - S_{diff}$ and $S_{west} - S_{diff}$ fluxes on September 18th showing that linear extrapolation of $S_{west} - S_{diff}$ can be used to determine Δ.

**S7. Comparison between $S_{west}$ and $S_{south}$ measured and modeled using Eq.5-7 with $S_{up}$ and $S_{down}$**

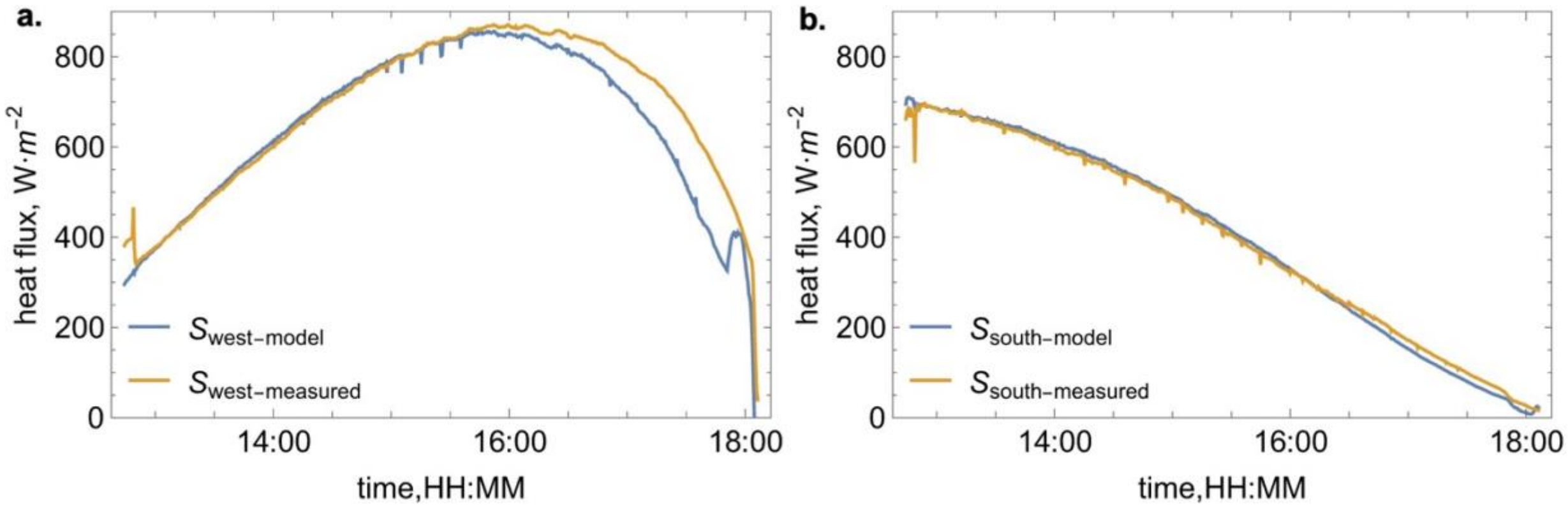

**Fig.S7** Comparison of **a.** $S_{west}$ and **b.** $S_{south}$ measured and modeled according to $S_{west} = \left((1+\rho)S_{dir}sin\theta + S_{diff}\right)sin\phi^*$ and $S_{south} = \left((1+\rho)S_{dir}sin\theta + S_{diff}\right)cos\phi^*$ with $S_{diff}, S_{dir}$, and $\rho$ calculated using Eq.5-7 with $S_{up}$ and $S_{down}$.

**S8. MaRTy and ANDI shortwave and longwave measurements in full shade**

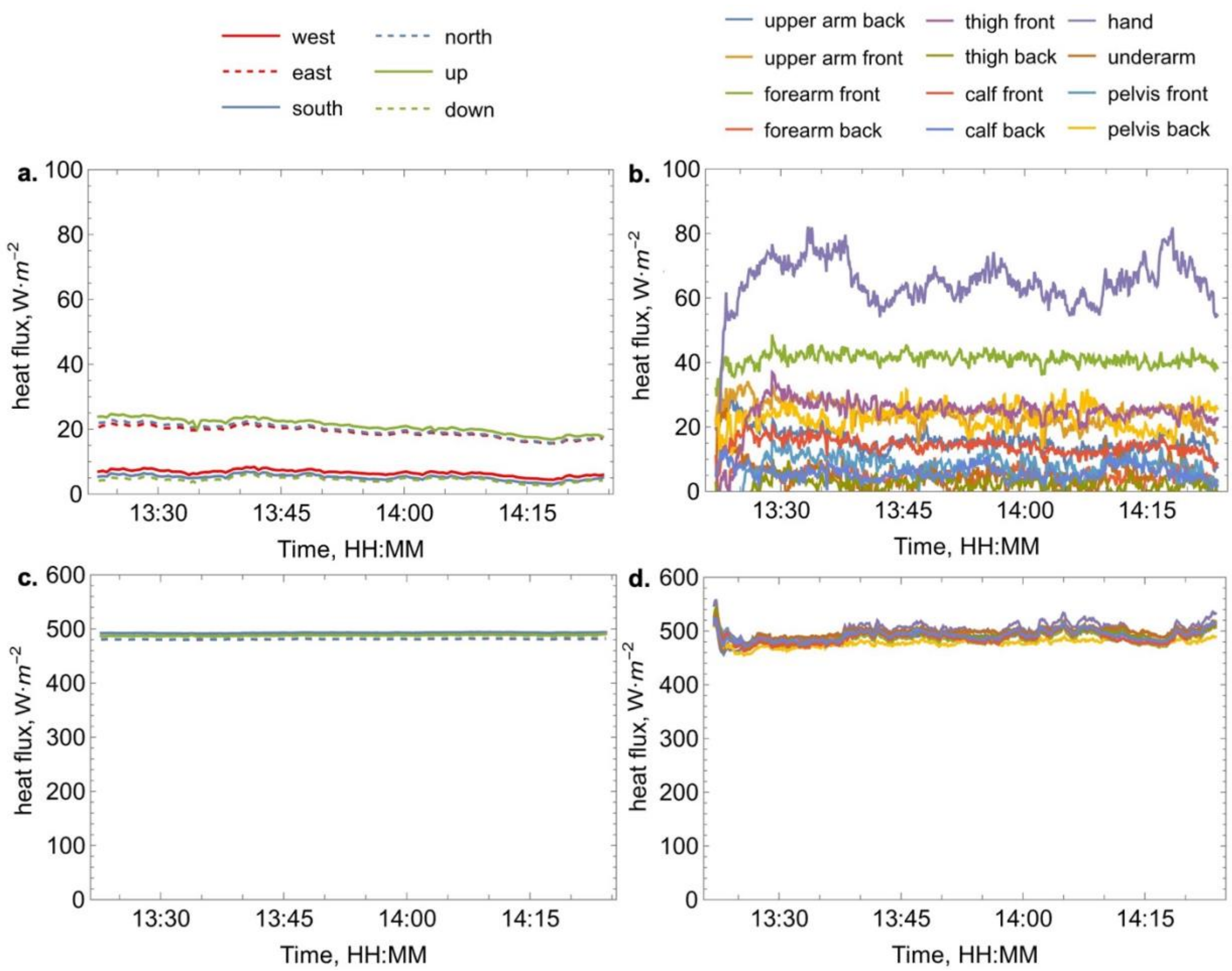

**Fig.S8 a.** MaRTy and **b.** ANDI shortwave and **c.** MaRTy and **d.** ANDI longwave measurements in full shade.

**S9. MaRTy and ANDI shortwave and longwave measurements in the narrow canyon**

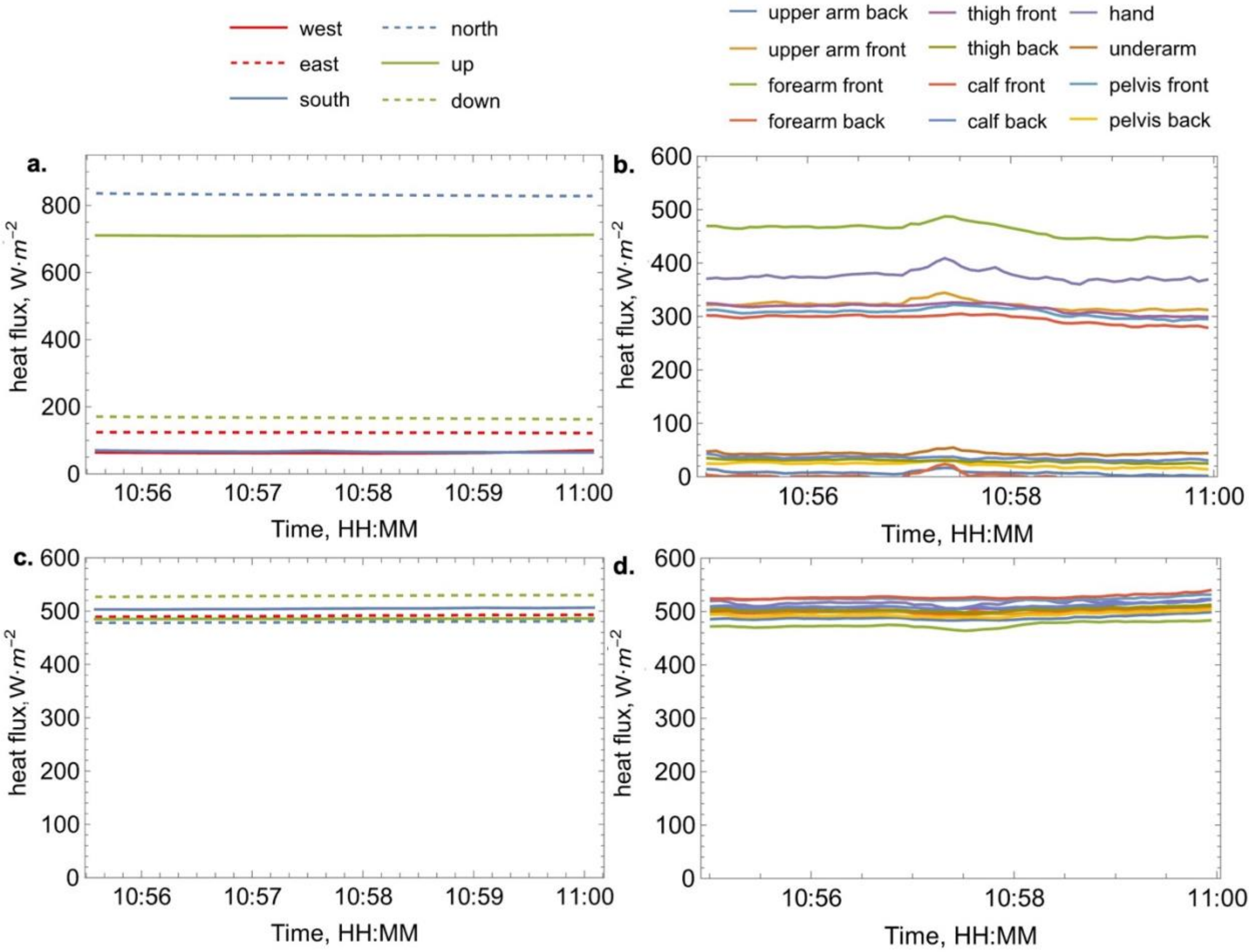

**Fig.S9 a.** MaRTy (note: the north facing sensor was oriented towards the sun during the measurement), and **b.** ANDI shortwave and **c.** MaRTy and **d.** ANDI longwave measurements in the canyon.

**S10. MaRTy and ANDI longwave measurements in partial shade**

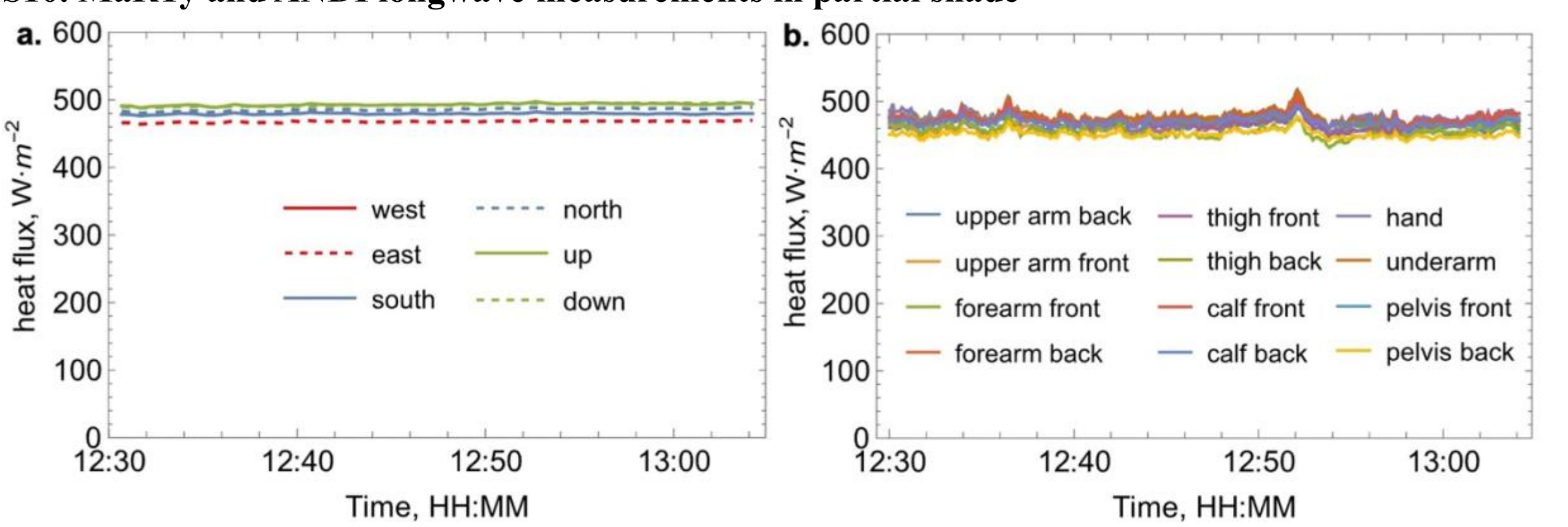

**Fig.S10 a.** MaRTy and **b.** ANDI longwave measurements in partial shade provided by Palo Verde tree.

## S11. Simulated and measured ANDI shortwave irradiation in partial shade

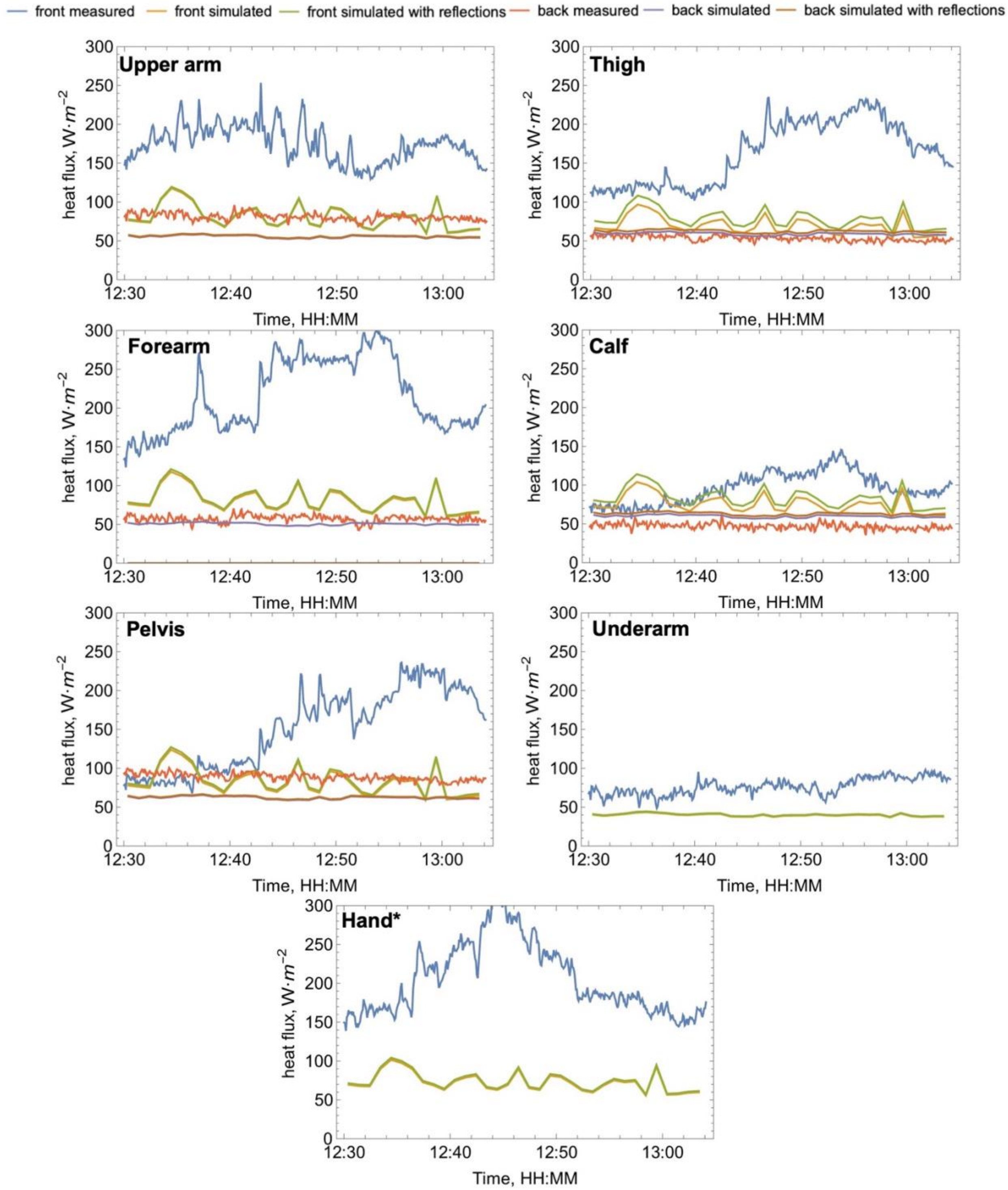

**Fig.S11** ANDI shortwave measurements in partial shade provided by Palo Verde tree. *Hand refers to the palm hand, see next Section.

**S12. The ANDI measured net heat fluxes with and without long sleeve shirts**

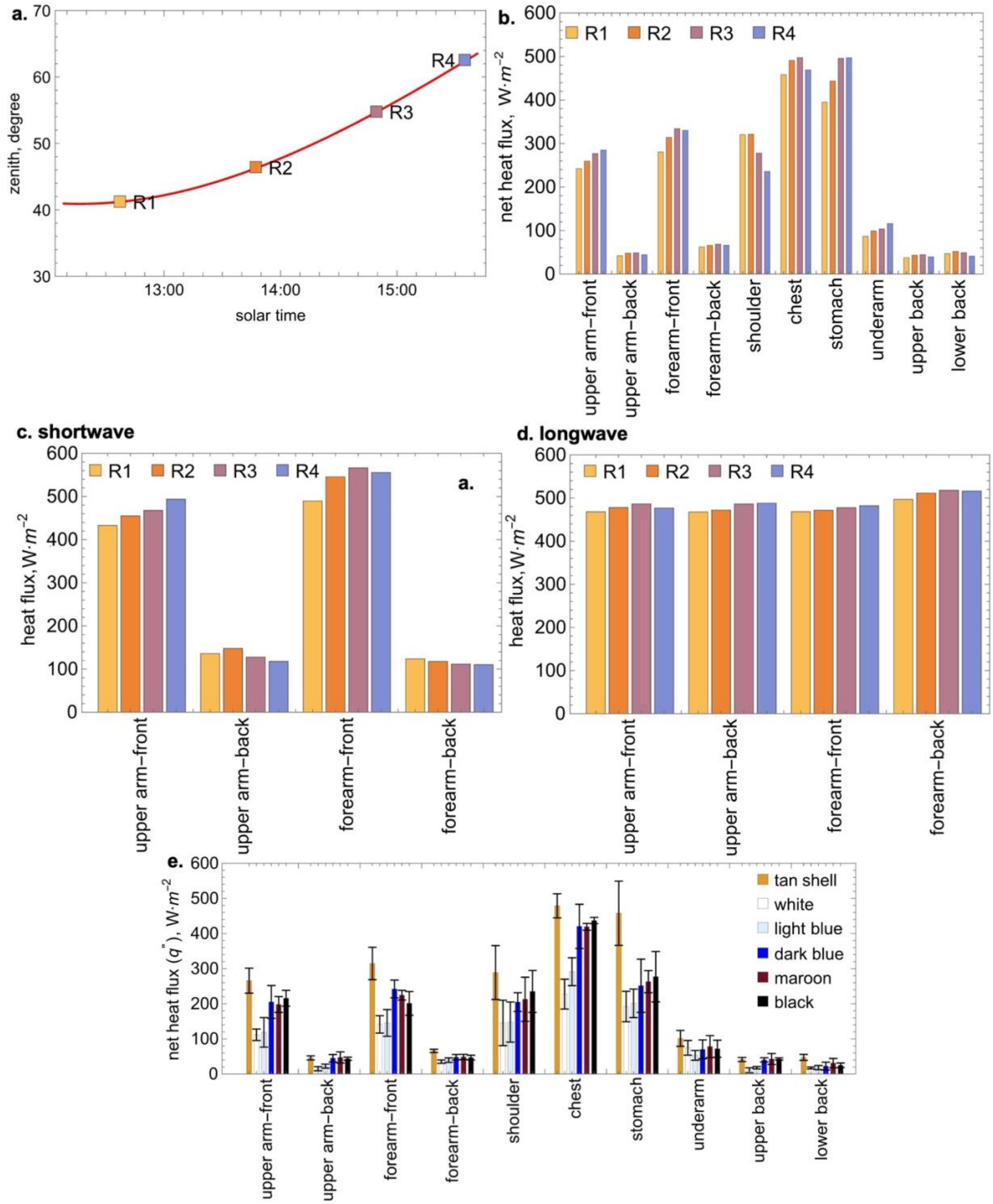

**Fig.S12 a.** the solar zenith angle with the four baseline "nude" ANDI repetitions (R1, R2, R3, R4) perform before, in between, and after measuring the shirt sets conducted on October 11th, 2024, **b.** net heat fluxes for each of the shirt covered zones compared across the repetitions, **and c.** shortwave and **d.** longwave heat fluxes measured during the four repetitions calculated using the white-tan zone method; e. average of the net heat fluxes across all the zones and repetitions (error bars indicate confidence interval of 80%).

**S13. Comparison of simulated radiation on hands and palms**

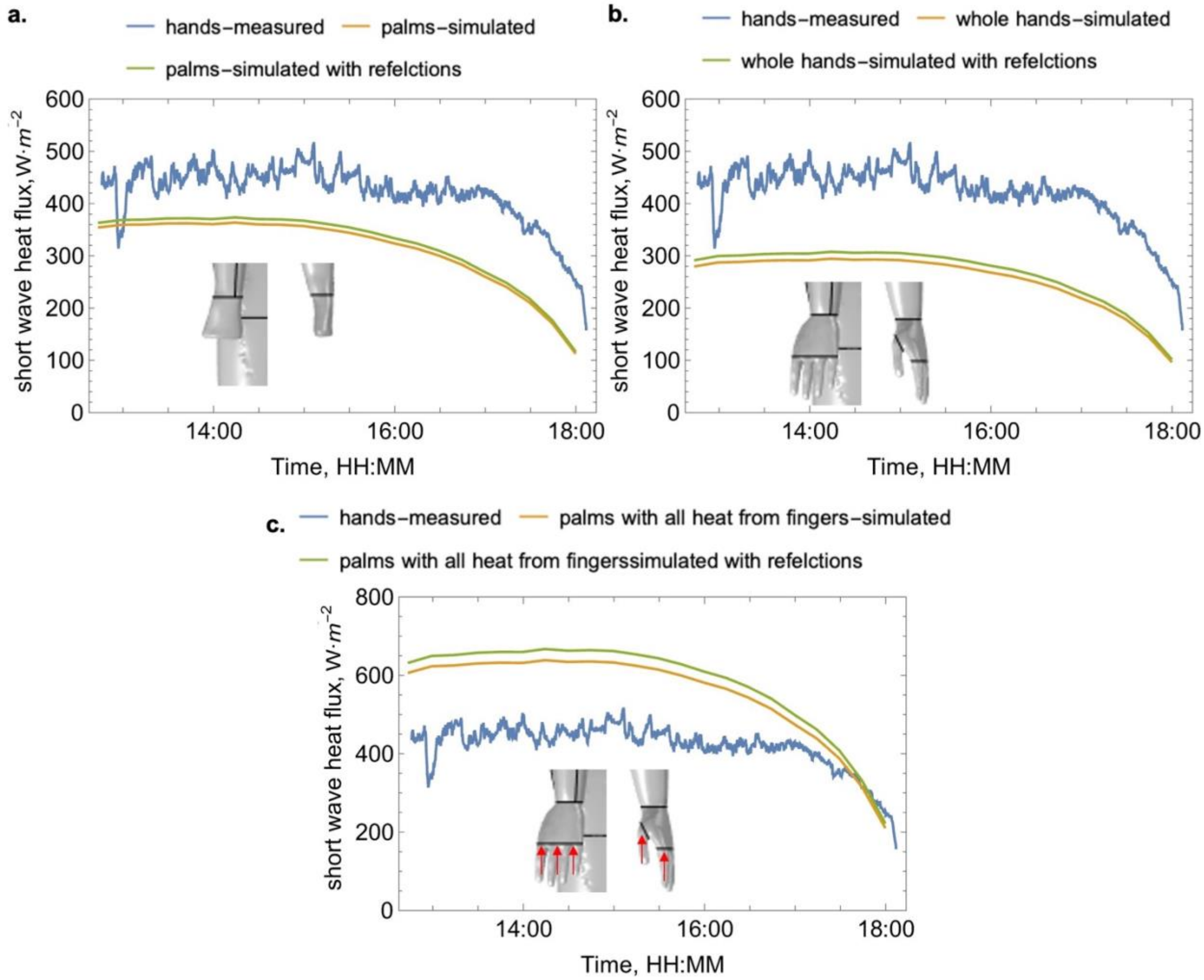

**Fig.S13** Comparison of ANDI-measured and simulated shortwave irradiation flux on hands modelled as **a.** just the palm (corresponding to actively cooled area with sensors on the manikin), **b.** hands with fingers (average flux is plotted), and **c.** flux on the palms with all the heat absorbed by fingers (average absorptivity for tan and white coating) being conducted to the palms.

**S14. Comparison of Mean Radiant Temperature (MRT) obtained from MaRTy fluxes using IRM method and virtual ANDI simulations**

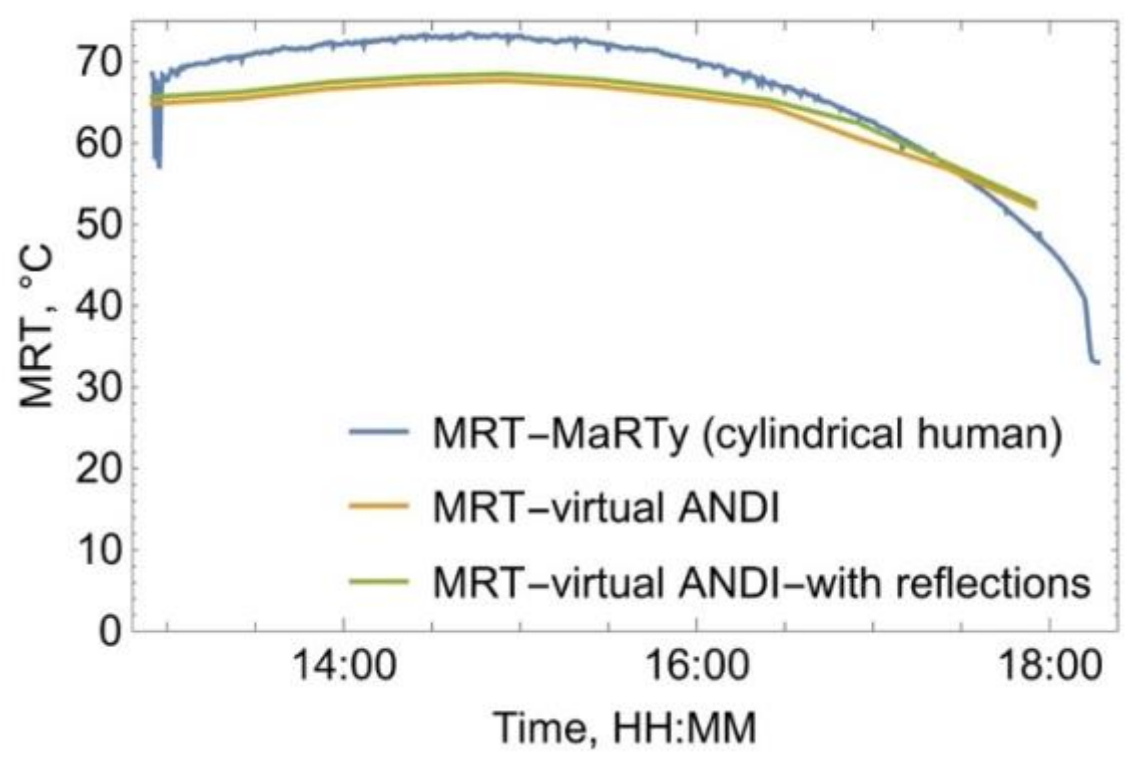

**Fig.S14** Comparison of MRT obtained from MaRTy shortwave and longwave fluxes measured on September 13th of 2024 using IRM methods (i.e., assuming cylindrical human, see Section 2.5) and using the virtual ANDI simulations with and without considering self-reflections on a uniformly tan-colored manikin.

**S15. Environmental conditions and ANDI shell temperature during the afternoon full sun experiments**

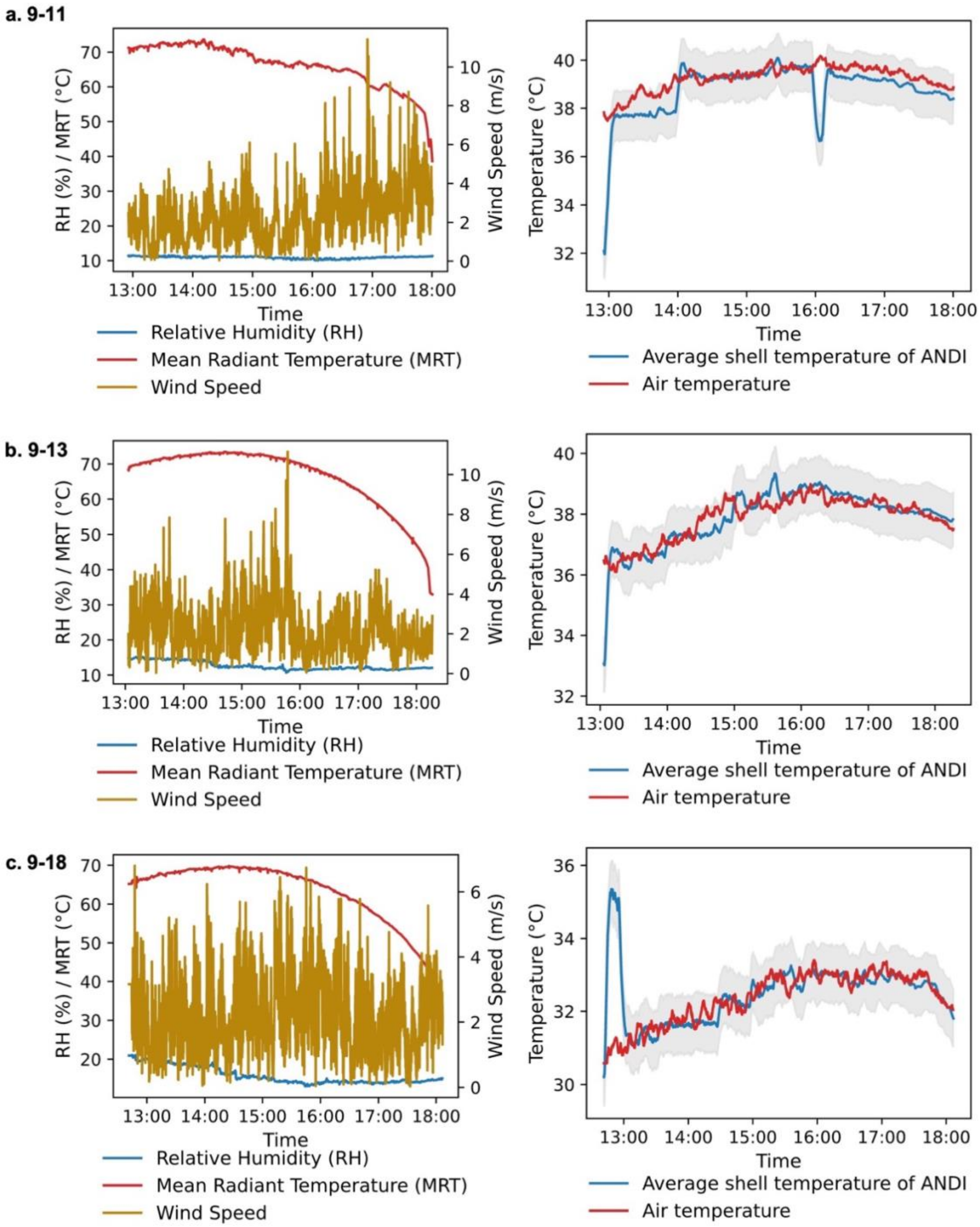

**Fig.S15** The relative humidity, mean radiant temperature (MaRTy IRM), wind speed, air temperature and average ANDI shell temperature (90% confidence interval shows variation across zones) during the experiments conducted on the afternoon of a. September 11$^{th}$, b. September 13$^{th}$, September 18$^{th}$ of 2024.

**References**

[1] K. Rykaczewski, A. Joshi, S.H. Viswanathan, S.S. Guddanti, K. Sadeghi, M. Gupta, A.K. Jaiswal, K. Kompally, G. Pathikonda, R. Barlett, J.K. Vanos, A. Middel, A simple three-cylinder radiometer and low-speed anemometer to characterize human extreme heat exposure, Int J Biometeorol 68 (2024) 1081–1092.

[2] ASTM F1291-16, Standard Test Method for Measuring the Thermal Insulation of Clothing Using a Heated Manikin, American Society for Testing and Materials Book of Standards 11 (2004) 1–7.

[3] B. Holmer, F. Lindberg, D. Rayner, S. Thorsson, How to transform the standing man from a box to a cylinder–a modified methodology to calculate mean radiant temperature in field studies and models, in: Proceedings of the 9th International Conference on Urban Climate (ICUC9), Toulouse, France, 2015: pp. 20–24.